# Causal Relay Networks

Ihn-Jung Baik and Sae-Young Chung

Department EE, KAIST, Daejeon, Korea,

Email: ihn_jung@kaist.ac.kr, sychung@ee.kaist.ac.kr

### Abstract

In this paper, we study causal discrete-memoryless relay networks (DMRNs). The network consists of multiple nodes, each of which can be a source, relay, and/or destination. In the network, there are two types of relays, i.e., relays with one sample delay (strictly causal) and relays without delay (causal) whose transmit signal depends not only on the past received symbols but also on the current received symbol. For this network, we derive two new cut-set bounds, one when the causal relays have their own messages and the other when not. Using examples of a causal vector Gaussian two-way relay channel and a causal vector Gaussian relay channel, we show that the new cut-set bounds can be achieved by a simple amplify-and-forward type relaying. Our result for the causal relay channel strengthens the previously known capacity result for the same channel by El Gamal, Hassanpour, and Mammen.

## I. Introduction

A model of relay networks was first introduced by van der Meulen in [1], [2]. In [3], Cover and El Gamal provided the cut-set bound for the relay channel and introduced two coding strategies, decode-and-forward and compress-and-forward. The coding strategies can achieve the capacities of some relay channels. For more complicated networks, a general cut-set bound is provided in [4]. In these relay networks, it is usually assumed that the relay's operation is strictly causal, i.e., its transmit symbol depends only on its past received symbols. Recently, new types of relays were introduced in [5], namely causal and noncausal relays. In the causal relay channel, the relay's transmit symbol depends on both the past and the current received symbols and in the noncausal relay channel, the relay's transmit symbol can depend on future received symbols as well. In [6], Wang and Naghshvar provided an improved upper bound for the noncausal relay channels. For the interference channel with a causal relay, Chang and Chung established outer bounds and also showed that they can be achieved if the relay's power exceeds a certain threshold and some additional conditions are satisfied [7]. In [8], cut-set bounds were established for DMRNs without ordering among casual relays. Also, we showed cut-set bounds for DMRNs with causal side information in [9]. Recently, Fong showed a cut-set bound for generalized networks including both strictly causal and causal relays in [10]. Also, Kramer developed a cut-set bound for a network with memory inside each block of symbols in [11].

In this paper, we focus on the DMRN with multiple sources and destinations. We assume each relay node is either causal or strictly causal. We show cut-set outer bounds for two types of networks, one with own messages at causal relays and the other without. Our bounds reduce to the classical cut-set bound [4] if there is no causal relay in the network. We provide examples of simple causal DMRN's such as the causal vector Gaussian two-way relay channel (TWRC) and the causal vector Gaussian relay channel and show a simple amplify-and-forward relaying can be optimal if the relay's power exceeds a certain threshold. Our result for the causal scalar Gaussian relay channel strengthens the previous capacity result for the same channel in [5] by removing a constraint in Proposition 9 in [5].

The remainder of this paper is organized as follows. In Section II, our network model is introduced. In Section III, we provide two new cut-set bounds for causal DMRN's and compare the bounds with previously known results and then, in Section IV, we give examples of causal DMRN's. Finally, we conclude the paper in Section V.

## II. MODEL

A multiple-source multiple-destination causal DMRN of $K$ nodes

$$\left( \mathcal{X}_1 \times \ldots \times \mathcal{X}_K, \prod_{j=1}^{|\mathcal{N}_0|} p(y_j | x_{\mathcal{N}_1}, x_{[1:j-1]}, y_{[1:j-1]}) p(y_{\mathcal{N}_1} | x_{\mathcal{N}_0}, x_{\mathcal{N}_1}, y_{\mathcal{N}_0}), \mathcal{Y}_1 \times \ldots \times \mathcal{Y}_K \right)$$

consists of sender alphabets $\mathcal{X}_k$ and receiver alphabets $\mathcal{Y}_k$, $k \in [1 : K] \triangleq \{1, 2, \ldots, K\}$ and a conditional pmf

$$\prod_{j=1}^{|\mathcal{N}_0|} p(y_j | x_{\mathcal{N}_1}, x_{[1:j-1]}, y_{[1:j-1]}) p(y_{\mathcal{N}_1} | x_{\mathcal{N}_0}, x_{\mathcal{N}_1}, y_{\mathcal{N}_0}), \tag{1}$$

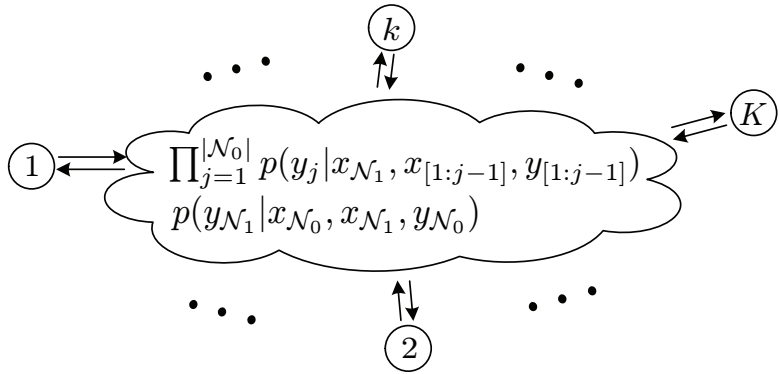

Fig. 1. Causal DMRN

where $\mathcal{N}_0 = \{1, 2, \ldots, |\mathcal{N}_0|\} \subseteq [1 : K]$ is the set of causal relays, and $\mathcal{N}_1 = [1 : K] \backslash \mathcal{N}_0$ is the set of strictly causal relays. Notations such as $x_{\mathcal{N}_0}$ denotes $\{x_k | k \in \mathcal{N}_0\}$. Note that the causal relays are ordered to avoid infinite loops among causal relays. A $\left(2^{nR_{11}}, 2^{nR_{12}}, \ldots, 2^{nR_{KK}}, n\right)$ code for this network consists of $K^2$ message sets, $K$ encoders, and $K$ decoders. Encoder $k \in \mathcal{N}_0$ assigns an input symbol $x_{ki}$ to each $(m_{k1}, \ldots, m_{kK}, y_k^i)$, and encoder $k \in \mathcal{N}_1$ assigns an input symbol $x_{ki}$ to each $(m_{k1}, \ldots, m_{kK}, y_k^{i-1})$ at time $i \in [1 : n]$, where $m_{kj}$ is the message from node $k$ to node $j$, and decoder $k \in [1 : K]$ assigns a set of message estimates $(\hat{m}_{1k}, \ldots, \hat{m}_{Kk})$ to each received sequence $y_k^n$, where $\hat{m}_{jk}$ is a decoded message from nodes $j$ to $k$. We assume $R_{kk} = 0$ for all $k \in [1 : K]$.

We also consider three special cases. The first is a causal DMRN without messages at causal relays, i.e., encoder $k \in \mathcal{N}_0$ assigns an input symbol $x_{ki}$ to each $y_k^i$ for $i \in [1 : n]$. The other two are causal DMRNs with/without messages at causal relays with a conditional pmf

$$p(y_{\mathcal{N}_0} | x_{\mathcal{N}_1}) p(y_{\mathcal{N}_1} | x_{\mathcal{N}_0}, x_{\mathcal{N}_1}, y_{\mathcal{N}_0}), \tag{2}$$

i.e., the received signals of causal relays depend on the transmit signals of the strictly causal relays only.

## III. MAIN RESULTS

In this section, we derive new cut-set bounds for the scenarios we consider. First, we show a cut-set bound for the causal DMRN, where each node in the network has its own messages, and then show a cut-set bound for the causal DMRN without messages at causal relays. For each scenario, we also consider the channel given as (2) as special cases.

*A. Causal DMRN with messages at causal relays (general case)*

We derive a cut-set bound for the causal DMRN when every node has messages to send.

*Theorem 1:* If the rates $R_{jk}$ are achievable in the causal DMRN with a conditional pmf given as (1), there exists some joint probability distribution

$$p(x_{\mathcal{N}_1}) \prod_{j=1}^{|\mathcal{N}_0|} \left\{ p(y_j | x_{\mathcal{N}_1}, x_{[1:j-1]}, y_{[1:j-1]}) p(x_j | x_{\mathcal{N}_1}, x_{[1:j-1]}, y_{[1:j]}) \right\} p(y_{\mathcal{N}_1} | x_{\mathcal{N}_0}, x_{\mathcal{N}_1}, y_{\mathcal{N}_0})$$

such that

$$
\begin{aligned}
\sum_{j \in \mathcal{S}, k \in \mathcal{S}^c} & R_{jk} \\
&\leq \sum_{j=1}^{|\mathcal{U}^c|} I \left( \begin{array}{c} X_{\mathcal{V}}, X_{[1:l_j-1] \setminus \{l_1, l_2, \ldots, l_{j-1}\}}, \\ Y_{[1:l_j-1] \setminus \{l_1, l_2, \ldots, l_{j-1}\}} \end{array} ; Y_{l_j} \left| \begin{array}{c} X_{\mathcal{V}^c}, X_{l_1}, X_{l_2}, \ldots, X_{l_{j-1}}, \\ Y_{l_1}, Y_{l_2}, \ldots, Y_{l_{j-1}} \end{array} \right. \right) \\
&\quad + I(X_{\mathcal{S}}, Y_{\mathcal{U}}; Y_{\mathcal{V}^c} | Y_{\mathcal{U}^c}, X_{\mathcal{S}^c}),
\end{aligned}
\tag{3}
$$

for all $\mathcal{S} \subset [1:K]$, where $\mathcal{U} = \mathcal{S} \cap \mathcal{N}_0$, $\mathcal{U}^c = \mathcal{N}_0 \setminus \mathcal{U} = \{l_1, \ldots, l_{|\mathcal{U}^c|}\}$, $l_i < l_j$ if $i < j$, $\mathcal{V} = \mathcal{S} \cap \mathcal{N}_1$, and $\mathcal{V}^c = \mathcal{N}_1 \setminus \mathcal{V}$.

*Proof:* See Appendix A. ∎

As a special case of a conditional pmf given as (2), we get the following cut-set bound.

*Corollary 1:* If the rates $R_{jk}$ are achievable in the causal DMRN with a conditional pmf given as (2), there exists some joint probability distribution

$$p(x_{\mathcal{N}_1}) \prod_{j=1}^{|\mathcal{N}_0|} \left\{ p(y_j | x_{\mathcal{N}_1}, y_{[1:j-1]}) p(x_j | x_{\mathcal{N}_1}, x_{[1:j-1]}, y_{[1:j]}) \right\} p(y_{\mathcal{N}_1} | x_{\mathcal{N}_0}, x_{\mathcal{N}_1}, y_{\mathcal{N}_0}) \tag{4}$$

such that

$$
\begin{aligned}
\sum_{j \in \mathcal{S}, k \in \mathcal{S}^c} R_{jk} &\leq \sum_{j=1}^{|\mathcal{U}^c|} I \left( X_{\mathcal{V}}, Y_{[1:l_j-1] \setminus \{l_1, \ldots, l_{j-1}\}}; Y_{l_j} | X_{\mathcal{V}^c}, Y_{l_1}, \ldots, Y_{l_{j-1}} \right) \\
&\quad + I(X_{\mathcal{S}}, Y_{\mathcal{U}}; Y_{\mathcal{V}^c} | Y_{\mathcal{U}^c}, X_{\mathcal{S}^c}),
\end{aligned}
\tag{5}
$$

for all $\mathcal{S} \subset [1:K]$, where $\mathcal{U} = \mathcal{S} \cap \mathcal{N}_0$, $\mathcal{U}^c = \mathcal{N}_0 \setminus \mathcal{U} = \{l_1, \ldots, l_{|\mathcal{U}^c|}\}$, $l_i < l_j$ if $i < j$, $\mathcal{V} = \mathcal{S} \cap \mathcal{N}_1$, and $\mathcal{V}^c = \mathcal{N}_1 \setminus \mathcal{V}$.

*Proof:* The cut-set bound (3) becomes

$$\sum_{j \in \mathcal{S}, k \in \mathcal{S}^c} R_{jk} \le \sum_{j=1}^{|\mathcal{U}^c|} I \left( \begin{array}{c} X_{\mathcal{V}}, X_{[1:l_j-1]\setminus\{l_1,\ldots,l_{j-1}\}}, \\ Y_{[1:l_j-1]\setminus\{l_1,\ldots,l_{j-1}\}} \end{array} ; Y_{l_j} \middle| \begin{array}{c} X_{\mathcal{V}^c}, X_{l_1}, \ldots, X_{l_{j-1}}, \\ Y_{l_1}, \ldots, Y_{l_{j-1}} \end{array} \right)$$
$$\quad + I(X_{\mathcal{S}}, Y_{\mathcal{U}}; Y_{\mathcal{V}^c} | Y_{\mathcal{U}^c}, X_{\mathcal{S}^c})$$

$$= \sum_{j=1}^{|\mathcal{U}^c|} \left\{ H \left( Y_{l_j} \middle| \begin{array}{c} X_{\mathcal{V}^c}, \\ X_{l_1}, \ldots, X_{l_{j-1}}, \\ Y_{l_1}, \ldots, Y_{l_{j-1}} \end{array} \right) - H \left( Y_{l_j} \middle| \begin{array}{c} X_{\mathcal{V}}, X_{\mathcal{V}^c}, \\ X_{[1:l_j-1]}, \\ Y_{[1:l_j-1]} \end{array} \right) \right\}$$
$$\quad + I(X_{\mathcal{S}}, Y_{\mathcal{U}}; Y_{\mathcal{V}^c} | Y_{\mathcal{U}^c}, X_{\mathcal{S}^c})$$

$$\overset{(a)}{\le} \sum_{j=1}^{|\mathcal{U}^c|} \left\{ H \left( Y_{l_j} | X_{\mathcal{V}^c}, Y_{l_1}, \ldots, Y_{l_{j-1}} \right) - H \left( Y_{l_j} \middle| \begin{array}{c} X_{\mathcal{V}}, X_{\mathcal{V}^c}, \\ X_{[1:l_j-1]}, Y_{[1:l_j-1]} \end{array} \right) \right\}$$
$$\quad + I(X_{\mathcal{S}}, Y_{\mathcal{U}}; Y_{\mathcal{V}^c} | Y_{\mathcal{U}^c}, X_{\mathcal{S}^c})$$

$$\overset{(b)}{=} \sum_{j=1}^{|\mathcal{U}^c|} \{ H(Y_{l_j} | X_{\mathcal{V}^c}, Y_{l_1}, \ldots, Y_{l_{j-1}}) - H(Y_{l_j} | X_{\mathcal{V}}, X_{\mathcal{V}^c}, Y_{[1:l_j-1]}) \}$$
$$\quad + I(X_{\mathcal{S}}, Y_{\mathcal{U}}; Y_{\mathcal{V}^c} | Y_{\mathcal{U}^c}, X_{\mathcal{S}^c})$$

$$= \sum_{j=1}^{|\mathcal{U}^c|} I \left( X_{\mathcal{V}}, Y_{[1:l_j-1]\setminus\{l_1,\ldots,l_{j-1}\}}; Y_{l_j} | X_{\mathcal{V}^c}, Y_{l_1}, \ldots, Y_{l_{j-1}} \right)$$
$$\quad + I(X_{\mathcal{S}}, Y_{\mathcal{U}}; Y_{\mathcal{V}^c} | Y_{\mathcal{U}^c}, X_{\mathcal{S}^c}),$$

where $(a)$ is because conditioning reduces entropy, $(b)$ is because of Markov chain $X_{[1:l_j-1]} \to (X_{\mathcal{V}}, X_{\mathcal{V}^c}, Y_{[1:l_j-1]}) \to Y_{l_j}$. Finally, we get the cut-set bound equivalent to the bound (5). ∎

*Proposition 1:* The cut-set bound in Corollary 1 coincides with the classical cut-set bound [4] if $\mathcal{N}_0 = \emptyset$.

*Proof:* If $\mathcal{N}_0 = \emptyset$, the probability distribution (4) becomes

$$p(x_{\mathcal{N}_1}) p(y_{\mathcal{N}_1} | x_{\mathcal{N}_1}).$$

Then, the mutual information in (5) becomes

$$\sum_{j \in \mathcal{S}, k \in \mathcal{S}^c} R_{jk} \leq I(X_{\mathcal{S}}; Y_{\mathcal{V}^c} | X_{\mathcal{S}^c})$$

$$= I(X_{\mathcal{S}}; Y_{\mathcal{S}^c} | X_{\mathcal{S}^c}),$$

which is same as the classical cut-set bound. ∎

### B. Causal DMRN without messages at causal relays

In this subsection, we consider the causal DMRN in which the causal relays do not have their own messages. Under the restriction, we show a tighter bound in this subsection than the one in the previous subsection.

*Theorem 2:* If the rates $R_{jk}$ are achievable in the causal DMRN without messages at causal relays with a conditional pmf given as (1), there exists some joint probability distribution

$$p(u_{\mathcal{N}_0}, x_{\mathcal{N}_1}) \prod_{j=1}^{|\mathcal{N}_0|} p(y_j | x_{\mathcal{N}_1}, x_{[1:j-1]}, y_{[1:j-1]}) p(y_{\mathcal{N}_1} | x_{\mathcal{N}_0}, x_{\mathcal{N}_1}, y_{\mathcal{N}_0}) \qquad (6)$$

and $x_k = x_k(y_k, u_k)$ for $k \in \mathcal{N}_0$ such that

$$\sum_{j \in \mathcal{V}, k \in \mathcal{S}^c} R_{jk} \leq \sum_{j=1}^{|\mathcal{U}^c|} I \left( \begin{array}{c} X_{\mathcal{V}}, X_{[1:l_j-1] \setminus \{l_1, \ldots, l_{j-1}\}}, \\ Y_{[1:l_j-1] \setminus \{l_1, \ldots, l_{j-1}\}} \end{array} ; Y_{l_j} \left| \begin{array}{c} U_{\mathcal{U}^c}, X_{\mathcal{V}^c}, \\ X_{l_1}, \ldots, X_{l_{j-1}}, \\ Y_{l_1}, \ldots, Y_{l_{j-1}} \end{array} \right. \right)$$

$$+ I(X_{\mathcal{V}}, U_{\mathcal{U}}; Y_{\mathcal{V}^c} | U_{\mathcal{U}^c}, Y_{\mathcal{U}^c}, X_{\mathcal{V}^c}), \qquad (7)$$

for all $\mathcal{S} \subset [1 : K]$, where $\mathcal{U} = \mathcal{S} \cap \mathcal{N}_0$, $\mathcal{U}^c = \mathcal{N}_0 \setminus \mathcal{U} = \{l_1, \ldots, l_{|\mathcal{U}^c|}\}$, $l_i < l_j$ if $i < j$, $\mathcal{V} = \mathcal{S} \cap \mathcal{N}_1$, and $\mathcal{V}^c = \mathcal{N}_1 \setminus \mathcal{V}$.

*Proof:* See Appendix B. ∎

As a special case of a conditional pmf given as (2), we get the following cut-set bound.

*Corollary 2:* If the rates $R_{jk}$ are achievable in the causal DMRN without messages at causal relays with a conditional pmf given as (2), there exists some joint probability distribution

$$p(u_{\mathcal{N}_0}, x_{\mathcal{N}_1}) \prod_{j=1}^{|\mathcal{N}_0|} p(y_j | x_{\mathcal{N}_1}, y_{[1:j-1]}) p(y_{\mathcal{N}_1} | x_{\mathcal{N}_0}, x_{\mathcal{N}_1}, y_{\mathcal{N}_0}) \qquad (8)$$

and $x_k = x_k(y_k, u_k)$ for $k \in \mathcal{N}_0$ such that

$$\sum_{j \in \mathcal{V}, k \in \mathcal{S}^c} R_{jk} \leq \sum_{j=1}^{|\mathcal{U}^c|} I(X_{\mathcal{V}}, Y_{[1:l_j-1] \setminus \{l_1, \ldots, l_{j-1}\}}; Y_{l_j} | U_{\mathcal{U}^c}, X_{\mathcal{V}^c}, Y_{l_1}, \ldots, Y_{l_{j-1}}) \tag{9}$$
$$+ I(X_{\mathcal{V}}, U_{\mathcal{U}}; Y_{\mathcal{V}^c} | U_{\mathcal{U}^c}, Y_{\mathcal{U}^c}, X_{\mathcal{V}^c}),$$

for all $\mathcal{S} \subset [1 : K]$, where $\mathcal{U} = \mathcal{S} \cap \mathcal{N}_0$, $\mathcal{U}^c = \mathcal{N}_0 \setminus \mathcal{U} = \{l_1, \ldots, l_{|\mathcal{U}^c|}\}$, $l_i < l_j$ if $i < j$, $\mathcal{V} = \mathcal{S} \cap \mathcal{N}_1$, and $\mathcal{V}^c = \mathcal{N}_1 \setminus \mathcal{V}$.

*Proof:* Because $y_{\mathcal{N}_0}$ depends only on $x_{\mathcal{N}_1}$, the cut-set bound (7) becomes

$$\sum_{j \in \mathcal{S}, k \in \mathcal{S}^c} R_{jk} \leq \sum_{j=1}^{|\mathcal{U}^c|} I \left( \begin{array}{c} X_{\mathcal{V}}, X_{[1:l_j-1] \setminus \{l_1, \ldots, l_{j-1}\}}, \\ Y_{[1:l_j-1] \setminus \{l_1, \ldots, l_{j-1}\}} \end{array} ; Y_{l_j} \left| \begin{array}{c} U_{\mathcal{U}^c}, X_{\mathcal{V}^c}, \\ X_{l_1}, \ldots, X_{l_{j-1}}, \\ Y_{l_1}, \ldots, Y_{l_{j-1}} \end{array} \right. \right)$$

$$+ I(X_{\mathcal{V}}, U_{\mathcal{U}}; Y_{\mathcal{V}^c} | U_{\mathcal{U}^c}, Y_{\mathcal{U}^c}, X_{\mathcal{V}^c})$$

$$= \sum_{j=1}^{|\mathcal{U}^c|} \left\{ H \left( Y_{l_j} \left| \begin{array}{c} U_{\mathcal{U}^c}, X_{\mathcal{V}^c}, \\ X_{l_1}, \ldots, X_{l_{j-1}}, \\ Y_{l_1}, \ldots, Y_{l_{j-1}} \end{array} \right. \right) - H \left( Y_{l_j} \left| \begin{array}{c} U_{\mathcal{U}^c}, X_{\mathcal{V}}, X_{\mathcal{V}^c}, \\ X_{[1:l_j-1]}, Y_{[1:l_j-1]} \end{array} \right. \right) \right\}$$

$$+ I(X_{\mathcal{V}}, U_{\mathcal{U}}; Y_{\mathcal{V}^c} | U_{\mathcal{U}^c}, Y_{\mathcal{U}^c}, X_{\mathcal{V}^c})$$

$$\overset{(a)}{\leq} \sum_{j=1}^{|\mathcal{U}^c|} \left\{ H \left( Y_{l_j} \left| \begin{array}{c} U_{\mathcal{U}^c}, X_{\mathcal{V}^c}, \\ Y_{l_1}, \ldots, Y_{l_{j-1}} \end{array} \right. \right) - H \left( Y_{l_j} \left| \begin{array}{c} U_{\mathcal{U}^c}, X_{\mathcal{V}}, X_{\mathcal{V}^c}, \\ X_{[1:l_j-1]}, Y_{[1:l_j-1]} \end{array} \right. \right) \right\}$$

$$+ I(X_{\mathcal{V}}, U_{\mathcal{U}}; Y_{\mathcal{V}^c} | U_{\mathcal{U}^c}, Y_{\mathcal{U}^c}, X_{\mathcal{V}^c})$$

$$\overset{(b)}{=} \sum_{j=1}^{|\mathcal{U}^c|} \{ H(Y_{l_j} | U_{\mathcal{U}^c}, X_{\mathcal{V}^c}, Y_{l_1}, \ldots, Y_{l_{j-1}}) - H(Y_{l_j} | U_{\mathcal{U}^c}, X_{\mathcal{V}}, X_{\mathcal{V}^c}, Y_{[1:l_j-1]}) \}$$

$$+ I(X_{\mathcal{V}}, U_{\mathcal{U}}; Y_{\mathcal{V}^c} | U_{\mathcal{U}^c}, Y_{\mathcal{U}^c}, X_{\mathcal{V}^c})$$

$$= \sum_{j=1}^{|\mathcal{U}^c|} I(X_{\mathcal{V}}, Y_{[1:l_j-1] \setminus \{l_1, \ldots, l_{j-1}\}}; Y_{l_j} | U_{\mathcal{U}^c}, X_{\mathcal{V}^c}, Y_{l_1}, \ldots, Y_{l_{j-1}})$$

$$+ I(X_{\mathcal{V}}, U_{\mathcal{U}}; Y_{\mathcal{V}^c} | U_{\mathcal{U}^c}, Y_{\mathcal{U}^c}, X_{\mathcal{V}^c})$$

where $(a)$ is because conditioning reduces entropy, $(b)$ is because of Markov chain $X_{[1:l_j-1]} \to (U_{\mathcal{U}^c}, X_{\mathcal{V}}, X_{\mathcal{V}^c}, Y_{[1:l_j-1]}) \to Y_{l_j}$. Finally, we get the cut-set bound equivalent to the bound (9). ∎

## C. Comparison of two cut-set bounds

*Proposition 2:* If there are no messages at the causal relays, the cut-set bound for the causal DMRN shown in Theorem 1 includes the cut-set bound for the causal DMRN without messages at causal relays shown in Theorem 2.

*Proof:* Assuming the probability distribution (6), we get the following inequality.

$$
\sum_{j=1}^{|\mathcal{U}^c|} I\left(
\begin{array}{c}
X_{\mathcal{V}}, X_{[1:l_j-1]\backslash\{l_1,\ldots,l_{j-1}\}}, \\
Y_{[1:l_j-1]\backslash\{l_1,\ldots,l_{j-1}\}}
\end{array}
\;; Y_{l_j}\;\middle|
\begin{array}{c}
U_{\mathcal{U}^c}, X_{\mathcal{V}^c}, \\
X_{l_1},\ldots,X_{l_{j-1}}, \\
Y_{l_1},\ldots,Y_{l_{j-1}}
\end{array}
\right)
$$

$$
+ I(X_{\mathcal{V}}, U_{\mathcal{U}}; Y_{\mathcal{V}^c} | U_{\mathcal{U}^c}, Y_{\mathcal{U}^c}, X_{\mathcal{V}^c})
$$

$$
= \sum_{j=1}^{|\mathcal{U}^c|} \left\{ H\left( Y_{l_j} \middle|
\begin{array}{c}
U_{\mathcal{U}^c}, X_{\mathcal{V}^c}, \\
X_{l_1},\ldots,X_{l_{j-1}}, \\
Y_{l_1},\ldots,Y_{l_{j-1}}
\end{array}
\right) - H\left( Y_{l_j} \middle|
\begin{array}{c}
U_{\mathcal{U}^c}, X_{\mathcal{V}^c}, X_{\mathcal{V}}, \\
X_{[1:l_j-1]}, \\
Y_{[1:l_j-1]}
\end{array}
\right) \right\}
$$

$$
+ H(Y_{\mathcal{V}^c} | U_{\mathcal{U}^c}, Y_{\mathcal{U}^c}, X_{\mathcal{V}^c}) - H(Y_{\mathcal{V}^c} | U_{\mathcal{U}^c}, Y_{\mathcal{U}^c}, X_{\mathcal{V}^c}, X_{\mathcal{V}}, U_{\mathcal{U}})
$$

$$
\stackrel{(a)}{=} \sum_{j=1}^{|\mathcal{U}^c|} \left\{ H\left( Y_{l_j} \middle|
\begin{array}{c}
U_{\mathcal{U}^c}, X_{\mathcal{V}^c}, \\
X_{l_1},\ldots,X_{l_{j-1}}, \\
Y_{l_1},\ldots,Y_{l_{j-1}}
\end{array}
\right) - H\left( Y_{l_j} \middle|
\begin{array}{c}
X_{\mathcal{V}^c}, X_{\mathcal{V}}, \\
X_{[1:l_j-1]}, \\
Y_{[1:l_j-1]}
\end{array}
\right) \right\}
$$

$$
+ H(Y_{\mathcal{V}^c} | U_{\mathcal{U}^c}, Y_{\mathcal{U}^c}, X_{\mathcal{V}^c}, X_{\mathcal{U}^c}) - H(Y_{\mathcal{V}^c} | U_{\mathcal{U}^c}, Y_{\mathcal{U}^c}, X_{\mathcal{S}^c}, X_{\mathcal{V}}, U_{\mathcal{U}})
$$

$$
\stackrel{(b)}{\leq} \sum_{j=1}^{|\mathcal{U}^c|} \left\{ H\left( Y_{l_j} \middle|
\begin{array}{c}
X_{\mathcal{V}^c}, \\
X_{l_1},\ldots,X_{l_{j-1}}, \\
Y_{l_1},\ldots,Y_{l_{j-1}}
\end{array}
\right) - H\left( Y_{l_j} \middle|
\begin{array}{c}
X_{\mathcal{V}^c}, X_{\mathcal{V}}, \\
X_{[1:l_j-1]}, \\
Y_{[1:l_j-1]}
\end{array}
\right) \right\}
$$

$$
+ H(Y_{\mathcal{V}^c} | Y_{\mathcal{U}^c}, X_{\mathcal{V}^c}, X_{\mathcal{U}^c}) - H(Y_{\mathcal{V}^c} | U_{\mathcal{U}^c}, Y_{\mathcal{U}^c}, X_{\mathcal{S}^c}, X_{\mathcal{V}}, U_{\mathcal{U}}, Y_{\mathcal{U}})
$$

$$
\stackrel{(c)}{=} \sum_{j=1}^{|\mathcal{U}^c|} \left\{ H\left( Y_{l_j} \middle|
\begin{array}{c}
X_{\mathcal{V}^c}, \\
X_{l_1},\ldots,X_{l_{j-1}}, \\
Y_{l_1},\ldots,Y_{l_{j-1}}
\end{array}
\right) - H\left( Y_{l_j} \middle|
\begin{array}{c}
X_{\mathcal{V}^c}, X_{\mathcal{V}}, \\
X_{[1:l_j-1]}, \\
Y_{[1:l_j-1]}
\end{array}
\right) \right\}
$$

$$
+ H(Y_{\mathcal{V}^c} | Y_{\mathcal{U}^c}, X_{\mathcal{V}^c}, X_{\mathcal{U}^c}) - H(Y_{\mathcal{V}^c} | U_{\mathcal{U}^c}, Y_{\mathcal{U}^c}, X_{\mathcal{S}^c}, X_{\mathcal{S}}, U_{\mathcal{U}}, Y_{\mathcal{U}})
$$

$$\overset{(d)}{=} \sum_{j=1}^{|\mathcal{U}^c|} \left\{ H\left( Y_{l_j} \;\middle|\; \begin{matrix} X_{\mathcal{V}^c}, \\ X_{l_1}, \ldots, X_{l_{j-1}}, \\ Y_{l_1}, \ldots, Y_{l_{j-1}} \end{matrix} \right) - H\left( Y_{l_j} \;\middle|\; \begin{matrix} X_{\mathcal{V}^c}, X_{\mathcal{V}}, \\ X_{[1:l_j-1]}, \\ Y_{[1:l_j-1]} \end{matrix} \right) \right\}$$

$$+ H(Y_{\mathcal{V}^c} | Y_{\mathcal{U}^c}, X_{\mathcal{V}^c}, X_{\mathcal{U}^c}) - H(Y_{\mathcal{V}^c} | Y_{\mathcal{U}^c}, X_{\mathcal{S}^c}, X_{\mathcal{S}}, Y_{\mathcal{U}})$$

$$= \sum_{j=1}^{|\mathcal{U}^c|} I\left( \begin{matrix} X_{\mathcal{V}}, X_{[1:l_j-1] \setminus \{l_1, l_2, \ldots, l_{j-1}\}}, \\ Y_{[1:l_j-1] \setminus \{l_1, l_2, \ldots, l_{j-1}\}} \end{matrix} \;; Y_{l_j} \;\middle|\; \begin{matrix} X_{\mathcal{V}^c}, \\ X_{l_1}, X_{l_2}, \ldots, X_{l_{j-1}}, \\ Y_{l_1}, Y_{l_2}, \ldots, Y_{l_{j-1}} \end{matrix} \right)$$

$$+ I(X_{\mathcal{S}}, Y_{\mathcal{U}}; Y_{\mathcal{V}^c} | Y_{\mathcal{U}^c}, X_{\mathcal{S}^c}),$$

where $(a)$ is because $X_{\mathcal{U}^c}$ is a function of $(U_{\mathcal{U}^c}, Y_{\mathcal{U}^c})$ and a Markov chain

$U_{\mathcal{U}^c} \rightarrow (X_{\mathcal{V}}, X_{\mathcal{V}^c}, X_{[1:l_j-1]}, Y_{[1:l_j-1]}) \rightarrow Y_{l_j}$, $(b)$ is because conditioning reduces entropy, $(c)$ is because $X_{\mathcal{U}}$ is a function of $(U_{\mathcal{U}}, Y_{\mathcal{U}})$, and $(d)$ is because of the Markov chain $(U_{\mathcal{U}^c}, U_{\mathcal{U}}) \rightarrow (Y_{\mathcal{U}^c}, X_{\mathcal{V}^c}, X_{\mathcal{V}}, X_{\mathcal{U}^c}, Y_{\mathcal{U}}, X_{\mathcal{U}}) \rightarrow Y_{\mathcal{V}^c}$. ∎

In some channels, the cut-set bound in Theorem 1 is strictly larger than the cut-set bound in Theorem 2. Consider the following three-node relay channel, where $\mathcal{N}_0 = \{2\}$ and $\mathcal{N}_1 = \{1, 3\}$.

$$Y_2 = (X_1 + Z_2) \bmod 4$$

$$Y_3 = Y_2 \bmod 2,$$

where $Z_2$ is $\text{Bern}\left(\frac{1}{2}\right)$ and is independent of $X_1$, and the cardinality of $X_1$ is 4. For this channel, Theorem 1 becomes

$$\max_{p(x_1)p(x_2|x_1, y_2)} \min\{I(X_1; Y_2) + I(X_1; Y_3 | X_2, Y_2), I(X_1, X_2, Y_2; Y_3)\}$$

$$= \max_{p(x_1)p(x_2|x_1, y_2)} \min\{H(Y_2) - 1, H(Y_3)\}$$

$$= 1,$$

and Theorem 2 becomes

$$\max_{p(u_2,x_1),x_2(u_2,y_2)} \min\{I(X_1;Y_2,Y_3|U_2), I(X_1,U_2;Y_3)\}$$

$$\leq \max_{p(u_2,x_1),x_2(u_2,y_2)} I(X_1,U_2;Y_3)$$

$$= \max_{p(u_2,x_1),x_2(u_2,y_2)} H(Y_3) - H(Y_3|X_1,U_2)$$

$$\overset{(a)}{=} \max_{p(u_2,x_1),x_2(u_2,y_2)} H(Y_3) - H(Z_2|X_1,U_2)$$

$$= \max_{p(u_2,x_1),x_2(u_2,y_2)} H(Y_3) - H(Z_2)$$

$$= \max_{p(u_2,x_1),x_2(u_2,y_2)} H(Y_3) - 1$$

$$= 0,$$

where $(a)$ is because $Y_3 = (X_1 + Z_2) \bmod 2$. Thus, the rate region in Theorem 1 can be strictly larger than that of Theorem 2. However, Theorem 1 can still be useful since it can be used to prove some capacity results as in Section IV.

## IV. EXAMPLES

In this section, we use a causal vector Gaussian TWRC and a causal vector Gaussian relay channel as examples of causal DMRN's. We apply the cut-set bounds in Theorems 1 and 2 for the channels and show the bounds can be tight under some conditions.

### A. Causal vector Gaussian TWRC

In a causal TWRC, nodes 1 and 3 exchange their messages with help of causal relay node 2. Let $X_k$ denote the transmit signal and $Y_k$ denote the received signal of node $k \in [1:3]$. For this channel, $\mathcal{N}_0 = \{2\}$ and $\mathcal{N}_1 = \{1,3\}$. Then, the cut-set bound in Theorem 1 reduces to

$$R_{13} \leq \min \left\{ \begin{array}{l} I(X_1;Y_2|X_3) + I(X_1;Y_3|X_2,X_3,Y_2), \\ I(X_1,X_2,Y_2;Y_3|X_3) \end{array} \right\}$$

$$R_{31} \leq \min \left\{ \begin{array}{l} I(X_3;Y_2|X_1) + I(X_3;Y_1|X_2,X_1,Y_2), \\ I(X_3,X_2,Y_2;Y_1|X_1) \end{array} \right\}, \tag{10}$$

for some $p(x_1,x_3)p(x_2|x_1,x_3,y_2)$.

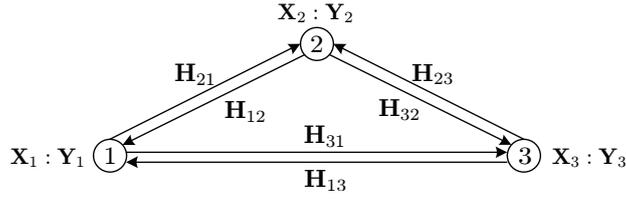

Fig. 2. Vector Gaussian TWRC with causal relay 2

For the causal vector Gaussian TWRC as shown in Fig. 2, the received signal at each node is

$$\mathbf{Y}_1 = \mathbf{H}_{12}\mathbf{X}_2 + \mathbf{H}_{13}\mathbf{X}_3 + \mathbf{Z}_1$$

$$\mathbf{Y}_2 = \mathbf{H}_{21}\mathbf{X}_1 + \mathbf{H}_{23}\mathbf{X}_3 + \mathbf{Z}_2$$

$$\mathbf{Y}_3 = \mathbf{H}_{32}\mathbf{X}_2 + \mathbf{H}_{31}\mathbf{X}_1 + \mathbf{Z}_3,$$

where $\mathbf{H}_{jk} \in \mathcal{C}^{r_j \times t_k}$ is the channel gain from nodes $k$ to $j$, $t_k$ is the number of transmit antennas of node $k$ and $r_j$ is the number of receive antennas of node $j$, $\mathrm{tr}(\mathbb{E}[\mathbf{X}_k\mathbf{X}_k^\dagger]) \leq P_k$ and $\mathbf{Z}_j \sim \mathcal{CN}(\mathbf{0}, \mathbf{I})$ for $k = 1, 2, 3$ and $j = 1, 2, 3$. Let $\Sigma_k = \mathbb{E}[\mathbf{X}_k\mathbf{X}_k^\dagger]$ for $k = 1, 2, 3$.

*Theorem 3:* If the transmit power of node 2 satisfies $\mathrm{tr}[\mathbf{F}(\mathbf{H}_{21}\Sigma_1\mathbf{H}_{21}^\dagger + \mathbf{H}_{23}\Sigma_3^*\mathbf{H}_{23}^\dagger + \mathbf{I})\mathbf{F}^\dagger] \leq P_2$, the capacity region of the causal vector Gaussian TWRC is

$$R_{13} \leq \max_{\mathrm{tr}(\Sigma_1) \leq P_1} \log |\Lambda_3\mathbf{V}_3^\dagger\Sigma_1\mathbf{V}_3\Lambda_3 + \mathbf{I}|$$

$$R_{31} \leq \max_{\mathrm{tr}(\Sigma_3) \leq P_3} \log |\Lambda_1\mathbf{V}_1^\dagger\Sigma_3\mathbf{V}_1\Lambda_1 + \mathbf{I}|,$$

where $\mathbf{F} = \begin{bmatrix} \mathbf{U}_{31}^\dagger\mathbf{H}_{32} \\ \mathbf{U}_{13}^\dagger\mathbf{H}_{12} \end{bmatrix}_r^{-1} \begin{bmatrix} \mathbf{U}_{21}^\dagger \\ \mathbf{U}_{23}^\dagger \end{bmatrix}$, $\Sigma_1^*$ and $\Sigma_3^*$ are chosen among values satisfying

$$\Sigma_1^* = \arg\max_{\mathrm{tr}(\Sigma_1) \leq P_1} \log |\Lambda_3\mathbf{V}_3^\dagger\Sigma_1\mathbf{V}_3\Lambda_3 + \mathbf{I}|$$

$$\Sigma_3^* = \arg\max_{\mathrm{tr}(\Sigma_3) \leq P_3} \log |\Lambda_1\mathbf{V}_1^\dagger\Sigma_3\mathbf{V}_1\Lambda_1 + \mathbf{I}|,$$

$\mathbf{U}_{21}, \mathbf{U}_{31}, \mathbf{U}_{23}, \mathbf{U}_{13}, \Lambda_1$, and $\Lambda_3$ are obtained by singular value decomposition (SVD) of channel matrices, i.e.,

$$\begin{bmatrix} \mathbf{H}_{21} \\ \mathbf{H}_{31} \end{bmatrix} = \begin{bmatrix} \mathbf{U}_{21} \\ \mathbf{U}_{31} \end{bmatrix} \Lambda_3\mathbf{V}_3^\dagger,$$

$$\begin{bmatrix} \mathbf{H}_{23} \\ \mathbf{H}_{13} \end{bmatrix} = \begin{bmatrix} \mathbf{U}_{23} \\ \mathbf{U}_{13} \end{bmatrix} \Lambda_1 \mathbf{V}_1^\dagger,$$

and $(\mathbf{A})_r^{-1} = \mathbf{A}^\dagger (\mathbf{A}\mathbf{A}^\dagger)^{-1}$ denotes the right inverse of $\mathbf{A}$. Here, $[\mathbf{U}_{21}^T \mathbf{U}_{31}^T]^T$, $[\mathbf{U}_{23}^T \mathbf{U}_{13}^T]^T$, $\mathbf{V}_1$, and $\mathbf{V}_3$ are unitary matrices, and $\Lambda_1$ and $\Lambda_3$ are diagonal matrices with positive entries.

*Proof:*

**Achievability.** Assume that $\mathbf{F}$ is chosen as in the theorem. Let nodes 1 and 3 transmit codewords using independent vector Gaussian codebooks with covariance matrices $\Sigma_1^*$ and $\Sigma_3^*$, respectively. We apply AF relaying using $\mathbf{F}$, i.e., the relay's transmit signal $\mathbf{X}_2$ is formed by $\mathbf{X}_2 = \mathbf{F}\,\mathbf{Y}_2$. Because $\mathbf{X}_2$ satisfies the power constraint from the condition given in the theorem, it can be transmitted from the relay. Then, the received signal at node 1 is

$$\begin{aligned} \mathbf{Y}_1 &= \mathbf{H}_{12}\,\mathbf{X}_2 + \mathbf{H}_{13}\,\mathbf{X}_3 + \mathbf{Z}_1 \\ &= \mathbf{H}_{12}\,\mathbf{F}\,\mathbf{Y}_2 + \mathbf{H}_{13}\,\mathbf{X}_3 + \mathbf{Z}_1 \\ &= \mathbf{H}_{12}\,\mathbf{F}(\mathbf{H}_{23}\,\mathbf{X}_3 + \mathbf{H}_{21}\,\mathbf{X}_1 + \mathbf{Z}_2) + \mathbf{H}_{13}\,\mathbf{X}_3 + \mathbf{Z}_1 \\ &= (\mathbf{H}_{12}\,\mathbf{F}\,\mathbf{H}_{23} + \mathbf{H}_{13})\,\mathbf{X}_3 + \mathbf{H}_{12}\,\mathbf{F}\,\mathbf{H}_{21}\,\mathbf{X}_1 + \mathbf{H}_{12}\,\mathbf{F}\,\mathbf{Z}_2 + \mathbf{Z}_1\,. \end{aligned}$$

Using the knowledge of $\mathbf{X}_1$ at node 1, the node can calculate $\tilde{\mathbf{Y}}_1$ defined as follows:

$$\begin{aligned} \tilde{\mathbf{Y}}_1 &\triangleq \mathbf{U}_{13}^\dagger (\mathbf{Y}_1 - \mathbf{H}_{12}\,\mathbf{F}\,\mathbf{H}_{21}\,\mathbf{X}_1) \\ &= \mathbf{U}_{13}^\dagger (\mathbf{H}_{12}\,\mathbf{F}\,\mathbf{U}_{23} + \mathbf{U}_{13})\Lambda_1 \mathbf{V}_1^\dagger\,\mathbf{X}_3 + \mathbf{U}_{13}^\dagger (\mathbf{H}_{12}\,\mathbf{F}\,\mathbf{Z}_2 + \mathbf{Z}_1) \\ &\overset{(a)}{=} (\mathbf{U}_{23}^\dagger \mathbf{U}_{23} + \mathbf{U}_{13}^\dagger \mathbf{U}_{13})\Lambda_1 \mathbf{V}_1^\dagger\,\mathbf{X}_3 + \mathbf{U}_{23}^\dagger \mathbf{Z}_2 + \mathbf{U}_{13}^\dagger \mathbf{Z}_1 \\ &\overset{(b)}{=} \Lambda_1 \mathbf{V}_1^\dagger\,\mathbf{X}_3 + \tilde{\mathbf{Z}}_1, \end{aligned}$$

where $(a)$ is because $\mathbf{U}_{13}^\dagger \mathbf{H}_{12} \mathbf{F} = \mathbf{U}_{23}^\dagger$ from the choice of $\mathbf{F} = \begin{bmatrix} \mathbf{U}_{31}^\dagger \mathbf{H}_{32} \\ \mathbf{U}_{13}^\dagger \mathbf{H}_{12} \end{bmatrix}^{-1} \begin{bmatrix} \mathbf{U}_{21}^\dagger \\ \mathbf{U}_{23}^\dagger \end{bmatrix}$, $(b)$ is because $[\mathbf{U}_{23}^T \mathbf{U}_{13}^T]^T$ is a unitary matrix, and $\tilde{\mathbf{Z}}_1 = \mathbf{U}_{23}^\dagger \mathbf{Z}_2 + \mathbf{U}_{13}^\dagger \mathbf{Z}_1$ with $\mathbb{E}[\tilde{\mathbf{Z}}_1 \tilde{\mathbf{Z}}_1^\dagger] = \mathbf{I}$. The covariance matrix of $\tilde{\mathbf{Y}}_1$ is

$$\mathbb{E}[\tilde{\mathbf{Y}}_1 \tilde{\mathbf{Y}}_1^\dagger] = \Lambda_1 \mathbf{V}_1^\dagger \Sigma_3^* \mathbf{V}_1 \Lambda_1 + \mathbf{I}.$$

Then, $R_{31}$ up to the following is achievable:

$$I(\mathbf{X}_3; \tilde{\mathbf{Y}}_1) = H(\tilde{\mathbf{Y}}_1) - H(\tilde{\mathbf{Y}}_1 \,|\, \mathbf{X}_3)$$
$$= H(\tilde{\mathbf{Y}}_1) - H(\tilde{\mathbf{Z}}_1)$$
$$= \log |\Lambda_1 \mathbf{V}_1^{\dagger} \Sigma_3^* \mathbf{V}_1 \Lambda_1 + \mathbf{I}|.$$

Similarly, $R_{13}$ up to the following is achievable:

$$I(\mathbf{X}_1; \tilde{\mathbf{Y}}_3) = \log |\Lambda_3 \mathbf{V}_3^{\dagger} \Sigma_1^* \mathbf{V}_3 \Lambda_3 + \mathbf{I}|.$$

**Converse.** Applying the cut-set bound (10) to the causal vector Gaussian TRWC, the first term in the bound for $R_{13}$ in (10) can be written as follows:

$$I(\mathbf{X}_1; \mathbf{Y}_2 \,|\, \mathbf{X}_3) + I(\mathbf{X}_1; \mathbf{Y}_3 \,|\, \mathbf{X}_2, \mathbf{X}_3, \mathbf{Y}_2)$$

$$= h(\mathbf{Y}_2 \,|\, \mathbf{X}_3) - h(\mathbf{Y}_2 \,|\, \mathbf{X}_3, \mathbf{X}_1)$$
$$+ h(\mathbf{Y}_3 \,|\, \mathbf{Y}_2, \mathbf{X}_3, \mathbf{X}_2) - h(\mathbf{Y}_3 \,|\, \mathbf{Y}_2, \mathbf{X}_3, \mathbf{X}_2, \mathbf{X}_1)$$

$$\overset{(a)}{\leq} h(\mathbf{Y}_2') - h(\mathbf{Y}_2' \,|\, \mathbf{X}_1) + h(\mathbf{Y}_3' \,|\, \mathbf{Y}_2', \mathbf{X}_3, \mathbf{X}_2)$$
$$- h(\mathbf{Y}_3' \,|\, \mathbf{Y}_2', \mathbf{X}_3, \mathbf{X}_2, \mathbf{X}_1)$$

$$\overset{(b)}{\leq} h(\mathbf{Y}_2') - h(\mathbf{Y}_2' \,|\, \mathbf{X}_1) + h(\mathbf{Y}_3' \,|\, \mathbf{Y}_2')$$
$$- h(\mathbf{Y}_3' \,|\, \mathbf{Y}_2', \mathbf{X}_3, \mathbf{X}_2, \mathbf{X}_1)$$

$$\overset{(c)}{=} h(\mathbf{Y}_2') - h(\mathbf{Y}_2' \,|\, \mathbf{X}_1) + h(\mathbf{Y}_3' \,|\, \mathbf{Y}_2') - h(\mathbf{Y}_3' \,|\, \mathbf{Y}_2', \mathbf{X}_1)$$

$$= I(\mathbf{X}_1; \mathbf{Y}_2', \mathbf{Y}_3')$$

$$= h(\mathbf{Y}_2', \mathbf{Y}_3') - h(\mathbf{Z}_2, \mathbf{Z}_3)$$

$$\leq \log(\pi e)^{(r_2 + r_3)} |K_{\hat{\mathbf{Y}}_3}| - \log(\pi e)^{(r_2 + r_3)}$$

$$= \log |K_{\hat{\mathbf{Y}}_3}|$$

$$= \log \left| \mathbb{E} \left[ \left[ \begin{array}{c} \mathbf{Y}'_2 \\ \mathbf{Y}'_3 \end{array} \right] [\mathbf{Y}_2'^{\dagger} \ \mathbf{Y}_3'^{\dagger}] \right] \right|$$

$$= \log \left| [\mathbf{H}_{21}^T \ \mathbf{H}_{31}^T]^T \Sigma_1 [\mathbf{H}_{21}^{\dagger} \ \mathbf{H}_{31}^{\dagger}] + \mathbf{I} \right|$$

$$= \log \left| [\mathbf{U}_{21}^T \mathbf{U}_{31}^T]^T \Lambda_3 \mathbf{V}_3^{\dagger} \Sigma_1 \mathbf{V}_3 \Lambda_3 [\mathbf{U}_{21}^{\dagger} \ \mathbf{U}_{31}^{\dagger}] + \mathbf{I} \right|$$

$$= \log \left| \Lambda_3 \mathbf{V}_3^{\dagger} \Sigma_1 \mathbf{V}_3 \Lambda_3 + \mathbf{I} \right|$$

$$\leq \max_{\mathrm{tr}(\Sigma_1) \leq P_1} \left| \Lambda_3 \mathbf{V}_3^{\dagger} \Sigma_1 \mathbf{V}_3 \Lambda_3 + \mathbf{I} \right|,$$

where $\mathbf{Y}'_2 = \mathbf{H}_{21} \mathbf{X}_1 + \mathbf{Z}_2$, $\mathbf{Y}'_3 = \mathbf{H}_{31} \mathbf{X}_1 + \mathbf{Z}_3$, $\hat{\mathbf{Y}}_3 = [\mathbf{Y}_2'^T \ \mathbf{Y}_3'^T]^T$, $(a)$ and $(b)$ are because conditioning reduces entropy, and $(c)$ is because of the Markov chain $(\mathbf{X}_2, \mathbf{X}_3) \to (\mathbf{X}_1, \mathbf{Y}'_2) \to \mathbf{Y}'_3$.

Similarly, we get an upper bound on $R_{31}$. Thus, the rates are upper bounded as

$$R_{13} \leq \max_{\mathrm{tr}(\Sigma_1) \leq P_1} \log |\Lambda_3 \mathbf{V}_3^{\dagger} \Sigma_1 \mathbf{V}_3 \Lambda_3 + \mathbf{I}|$$

$$R_{31} \leq \max_{\mathrm{tr}(\Sigma_3) \leq P_3} \log |\Lambda_1 \mathbf{V}_1^{\dagger} \Sigma_3 \mathbf{V}_1 \Lambda_1 + \mathbf{I}|.$$

Since, the achievable rate region and the outer bound coincide, the proof is completed. ∎

As a simpler example of a causal vector Gaussian TWRC, assume $r_k = 1$ for $k = 1, 2, 3$, $t_1 = t_3 = 1$, and $t_2 = 2$. The received signal at each node is

$$Y_1 = h_{12} X_{21} + h_{13} X_3 + Z_1$$

$$Y_2 = h_{21} X_1 + h_{23} X_3 + Z_2$$

$$Y_3 = h_{32} X_{23} + h_{31} X_1 + Z_3,$$

where $h_{jk}$ is the channel gain from nodes $k$ to $j$, $\mathbf{H}_{12} = [h_{12} \ 0]$, $\mathbf{H}_{32} = [0 \ h_{32}]$, and $\mathbf{X}_2 = [X_{21} \ X_{23}]^T$. If $P_2 \geq \left( \frac{|h_{23}|^2}{|h_{12}|^2 |h_{13}|^2} + \frac{|h_{21}|^2}{|h_{32}|^2 |h_{31}|^2} \right) (|h_{21}|^2 P_1 + |h_{23}|^2 P_3 + 1)$, then the capacity can be achieved by choosing $\mathbf{F} = \left[ \frac{h_{23}^{\dagger}}{h_{12} h_{13}^{\dagger}} \ \frac{h_{21}^{\dagger}}{h_{32} h_{31}^{\dagger}} \right]^T$.

### B. Causal vector Gaussian relay channel

In this subsection, we consider a causal relay channel, where node 1 transmit its message to node 3 with help of a causal relay node 2, as a special case of a causal TWRC. For this channel, the cut-set bound of Theorem 2 reduces to

$$R_{13} \leq \max \min\{I(X_1; Y_2, Y_3 | U_2), I(X_1, U_2; Y_3)\}, \tag{11}$$

where the maximization is over $p(u_2, x_1)$ and $x_2 = x_2(u_2, y_2)$. This recovers the upper bound in Theorem 2 in [5]. If we use Theorem 1 instead, then we get a potentially looser bound given as

$$R_{13} \leq \max_{p(x_1)p(x_2|x_1,y_2)} \min \left\{ \begin{array}{l} I(X_1; Y_2) + I(X_1; Y_3 | X_2, Y_2), \\ I(X_1, X_2, Y_2; Y_3) \end{array} \right\}. \tag{12}$$

Note that this is the same as the result in Theorem 1 of [5] (if modified, see the footnote below), where they showed an upper bound for the non-causal relay channel with three nodes.[1]

For the causal vector Gaussian relay channel, the received signal at each node is

$$\mathbf{Y}_2 = \mathbf{H}_{21}\,\mathbf{X}_1 + \mathbf{Z}_2$$

$$\mathbf{Y}_3 = \mathbf{H}_{31}\,\mathbf{X}_1 + \mathbf{H}_{32}\,\mathbf{X}_2 + \mathbf{Z}_3,$$

where $\mathbf{H}_{jk} \in \mathcal{C}^{r_j \times t_k}$ is the channel gain from nodes $k$ to $j$, $t_k$ is the number of transmit antennas of node $k$ and $r_j$ is the number of receive antennas of node $j$, $\mathrm{tr}(\mathbb{E}[\mathbf{X}_k\mathbf{X}_k^{\dagger}]) \leq P_k$ and $\mathbf{Z}_j \sim \mathcal{CN}(\mathbf{0}, \mathbf{I})$ for $k = 1, 2$ and $j = 2, 3$. Let $\Sigma_k = \mathbb{E}[\mathbf{X}_k\mathbf{X}_k^{\dagger}]$ for $k = 1, 2$. For this channel, we can directly obtain the following from Theorem 3.

*Proposition 3:* If the transmit power of node 2 satisfies $\mathrm{tr}[\mathbf{F}(\mathbf{H}_{21}\,\Sigma_1^*\,\mathbf{H}_{21}^{\dagger} + \mathbf{I})\mathbf{F}^{\dagger}] \leq P_2$, the capacity region of the causal vector Gaussian relay channel is

$$R_{13} \leq \max_{\mathrm{tr}(\Sigma_1) \leq P_1} \log |\Lambda\mathbf{V}^{\dagger}\Sigma_1\mathbf{V}\Lambda + \mathbf{I}|,$$

where $\mathbf{F} = (\mathbf{U}_{31}^{\dagger}\mathbf{H}_{32})_r^{-1}\mathbf{U}_{21}^{\dagger}$, $\Sigma_1^*$ is chosen among values satisfying

$$\Sigma_1^* = \arg\max_{\mathrm{tr}(\Sigma_1) \leq P_1} \log |\Lambda\mathbf{V}^{\dagger}\Sigma_1\mathbf{V}\Lambda + \mathbf{I}|,$$

$\mathbf{U}_{21}, \mathbf{U}_{31}, \Lambda$ and $\mathbf{V}$ are obtained by singular value decomposition (SVD) of channel matrices, i.e.,

$$\begin{bmatrix} \mathbf{H}_{21} \\ \mathbf{H}_{31} \end{bmatrix} = \begin{bmatrix} \mathbf{U}_{21} \\ \mathbf{U}_{31} \end{bmatrix} \Lambda\mathbf{V}^{\dagger}.$$

Here, $[\mathbf{U}_{21}^T\mathbf{U}_{31}^T]^T$, and $\mathbf{V}$ are unitary matrices, and $\Lambda$ is diagonal matrix with positive entries.

---

[1] Note that the supremum in (7) in [5] should be over $p(x_1)p(x_2|x_1, y_2)$ not over $p(x_1, x_2)$. Furthermore, $I(X_1, X_2; Y_3)$ in (7) in [5] does not hold since $(M, Y_3^{i-1}) \rightarrow (X_{1i}, X_{2i}) \rightarrow Y_{3i}$ is not a Markov chain due to unlimited lookahead at the relay. If $I(X_1, X_2; Y_3)$ in (7) in [5] is replaced by $I(X_1, X_2, Y_2; Y_3)$, then it becomes a valid upper bound.

As a simpler example, consider a single-antenna channel given as

$$Y_2 = h_{21}X_1 + Z_2$$

$$Y_3 = h_{31}X_1 + h_{32}X_2 + Z_3.$$

For this causal single-antenna Gaussian relay channel, if

$$P_2 \geq \left( \frac{|h_{21}|^2}{|h_{31}|^2|h_{32}|^2} \right)(|h_{21}|^2 P_1 + 1), \tag{13}$$

then its capacity is given by $\log\{1 + (|h_{21}|^2 + |h_{31}|^2)P_1\}$. Note that the above capacity is achieved by AF with $\mathbf{F} = \frac{h_{21}^\dagger}{h_{32}h_{31}^\dagger}$. In Proposition 9 in [5], they showed a similar result as above, but our result is stronger since they require an additional condition $|h_{21}| \leq |h_{32}|$ in addition to (13).

## V. Conclusion

In this paper, we studied a causal discrete memoryless relay network consisting of both causal and strictly causal relays. In an analog relay system such as a full duplex relay, if the delay spread including the path through relay is much smaller than the inverse of the bandwidth, it can be modeled as a causal relay. For this channel, we presented two cut-set bounds, where one with messages at causal relays and the other without messages at causal relays. Because we considered a general channel model, our bounds cover many known results such as the classical cut-set bound. Also, surprisingly, the obtained outer bound can be achieved in some Gaussian channels based on AF relaying scheme under some conditions.

APPENDIX A

PROOF OF THEOREM 1

For some $\epsilon_n \to 0$ as $n \to \infty$, we get

$$n \sum_{j \in \mathcal{S}, k \in \mathcal{S}^c} R_{jk}$$

$$= \sum_{j \in \mathcal{S}, k \in \mathcal{S}^c} H(M_{jk})$$

$$= H(M_{\mathcal{T}})$$

$$\overset{(a)}{=} H(M_{\mathcal{T}}|M_{\mathcal{T}^c})$$

$$= I(M_{\mathcal{T}}; Y_{\mathcal{S}^c}^n | M_{\mathcal{T}^c}) + H(M_{\mathcal{T}}|Y_{\mathcal{S}^c}^n, M_{\mathcal{T}^c})$$

$$\overset{(b)}{\leq} I(M_{\mathcal{T}}; Y_{\mathcal{S}^c}^n | M_{\mathcal{T}^c}) + n\epsilon_n$$

$$= \sum_{i=1}^{n} I(M_{\mathcal{T}}; Y_{\mathcal{S}^c,i} | M_{\mathcal{T}^c}, Y_{\mathcal{S}^c}^{i-1}) + n\epsilon_n$$

$$= \sum_{i=1}^{n} \big[ H(Y_{\mathcal{S}^c,i} | M_{\mathcal{T}^c}, Y_{\mathcal{S}^c}^{i-1})$$

$$- H(Y_{\mathcal{S}^c,i} | \mathcal{M}, Y_{\mathcal{S}^c}^{i-1}) \big] + n\epsilon_n$$

$$\overset{(c)}{=} \sum_{i=1}^{n} \big[ H(Y_{\mathcal{U}^c,i} | M_{\mathcal{T}^c}, Y_{\mathcal{S}^c}^{i-1})$$

$$- H(Y_{\mathcal{U}^c,i} | \mathcal{M}, Y_{\mathcal{S}^c}^{i-1})$$

$$+ H(Y_{\mathcal{V}^c,i} | M_{\mathcal{T}^c}, Y_{\mathcal{S}^c}^{i-1}, Y_{\mathcal{U}^c,i})$$

$$- H(Y_{\mathcal{V}^c,i} | \mathcal{M}, Y_{\mathcal{S}^c}^{i-1}, Y_{\mathcal{U}^c,i}) \big] + n\epsilon_n$$

$$\overset{(d)}{=} \sum_{i=1}^{n} \big[ H(Y_{\mathcal{U}^c,i} | M_{\mathcal{T}^c}, Y_{\mathcal{S}^c}^{i-1}, X_{\mathcal{V}^c,i})$$

$$- H(Y_{\mathcal{U}^c,i} | \mathcal{M}, Y_{\mathcal{S}^c}^{i-1}, X_{\mathcal{V}^c,i})$$

$$+ H(Y_{\mathcal{V}^c,i} | M_{\mathcal{T}^c}, Y_{\mathcal{S}^c}^{i-1}, Y_{\mathcal{U}^c,i}, X_{\mathcal{S}^c,i})$$

$$- H(Y_{\mathcal{V}^c,i} | \mathcal{M}, Y_{\mathcal{S}^c}^{i-1}, Y_{\mathcal{U}^c,i}, X_{\mathcal{S}^c,i}) \big] + n\epsilon_n$$

$$\stackrel{(e)}{=} \sum_{i=1}^{n} \left[ \sum_{j=1}^{|\mathcal{U}^c|} \left\{ H(Y_{l_j,i} | M_{\mathcal{T}^c}, Y_{\mathcal{S}^c}^{i-1}, X_{\mathcal{V}^c,i}, Y_{l_1,i}, Y_{l_2,i}, \ldots, Y_{l_{j-1},i}, X_{l_1,i}, X_{l_2,i}, \ldots, X_{l_{j-1},i}) \right. \right.$$

$$- H(Y_{l_j,i} | \mathcal{M}, Y_{\mathcal{S}^c}^{i-1}, X_{\mathcal{V}^c,i}, Y_{l_1,i}, Y_{l_2,i}, \ldots, Y_{l_{j-1},i}, X_{l_1,i}, X_{l_2,i} \ldots, X_{l_{j-1},i}) \Big\}$$

$$+ H(Y_{\mathcal{V}^c,i} | M_{\mathcal{T}^c}, Y_{\mathcal{S}^c}^{i-1}, Y_{\mathcal{U}^c,i}, X_{\mathcal{S}^c,i})$$

$$\left. - H(Y_{\mathcal{V}^c,i} | \mathcal{M}, Y_{\mathcal{S}^c}^{i-1}, Y_{\mathcal{U}^c,i}, X_{\mathcal{S}^c,i}) \right] + n\epsilon_n$$

$$\stackrel{(f)}{\leq} \sum_{i=1}^{n} \left[ \sum_{j=1}^{|\mathcal{U}^c|} \left\{ H(Y_{l_j,i} | X_{\mathcal{V}^c,i}, Y_{l_1,i}, Y_{l_2,i}, \ldots, Y_{l_{j-1},i}, X_{l_1,i}, X_{l_2,i}, \ldots, X_{l_{j-1},i}) \right. \right.$$

$$- H(Y_{l_j,i} | \mathcal{M}, Y_{\mathcal{S}^c}^{i-1}, X_{\mathcal{V}^c,i}, Y_{[1:l_j-1],i}, X_{[1:l_j-1],i}, X_{\mathcal{V},i}) \Big\}$$

$$+ H(Y_{\mathcal{V}^c,i} | Y_{\mathcal{U}^c,i}, X_{\mathcal{S}^c,i})$$

$$\left. - H(Y_{\mathcal{V}^c,i} | \mathcal{M}, Y_{\mathcal{S}^c}^{i-1}, Y_{\mathcal{U}^c,i}, X_{\mathcal{S}^c,i}, X_{\mathcal{S},i}, Y_{\mathcal{U},i}) \right] + n\epsilon_n$$

$$\stackrel{(g)}{=} \sum_{i=1}^{n} \left[ \sum_{j=1}^{|\mathcal{U}^c|} \left\{ H(Y_{l_j,i} | X_{\mathcal{V}^c,i}, Y_{l_1,i}, Y_{l_2,i} \ldots, Y_{l_{j-1},i}, X_{l_1,i}, X_{l_2,i}, \ldots, X_{l_{j-1},i}) \right. \right.$$

$$- H(Y_{l_j,i} | X_{\mathcal{V}^c,i}, Y_{1[1:l_j-1],i}, X_{[1:l_j-1],i}, X_{\mathcal{V},i}) \Big\}$$

$$+ H(Y_{\mathcal{V}^c,i} | Y_{\mathcal{U}^c,i}, X_{\mathcal{S}^c,i})$$

$$\left. - H(Y_{\mathcal{V}^c,i} | Y_{\mathcal{U}^c,i}, X_{\mathcal{S}^c,i}, X_{\mathcal{S},i}, Y_{\mathcal{U},i}) \right] + n\epsilon_n$$

$$= \sum_{i=1}^{n} \left[ \sum_{j=1}^{|\mathcal{U}^c|} I \left( \begin{array}{c} X_{\mathcal{V},i}, X_{[1:l_j-1] \setminus \{l_1,l_2,\ldots,l_{j-1}\},i}, \\ Y_{[1:l_j-1] \setminus \{l_1,l_2,\ldots,l_{j-1}\},i} \end{array} ; Y_{l_j,i} \left| \begin{array}{c} X_{\mathcal{V}^c,i}, \\ X_{l_1,i}, X_{l_2,i}, \ldots, X_{l_{j-1},i}, \\ Y_{l_1,i}, Y_{l_2,i}, \ldots, Y_{l_{j-1},i} \end{array} \right. \right) \right.$$

$$\left. + I(X_{\mathcal{S},i}, Y_{\mathcal{U},i}; Y_{\mathcal{V}^c,i} | Y_{\mathcal{U}^c,i}, X_{\mathcal{S}^c,i}) \right] + n\epsilon_n$$

$$= \sum_{i=1}^{n} \left[ \sum_{j=1}^{|\mathcal{U}^c|} I \left( \begin{array}{c} X_{\mathcal{V},i}, X_{[1:l_j-1] \setminus \{l_1,l_2,\ldots,l_{j-1}\},i}, \\ Y_{[1:l_j-1] \setminus \{l_1,\ldots,l_{j-1}\},i} \end{array} ; Y_{l_j,i} \left| \begin{array}{c} X_{\mathcal{V}^c,i}, Q = i \\ X_{l_1,i}, X_{l_2,i}, \ldots, X_{l_{j-1},i}, \\ Y_{l_1,i}, \ldots, Y_{l_{j-1},i} \end{array} \right. \right) \right.$$

$$\left. + I(X_{\mathcal{S},i}, Y_{\mathcal{U},i}; Y_{\mathcal{V}^c,i} | Y_{\mathcal{U}^c,i}, X_{\mathcal{S}^c,i}, Q = i) \right] + n\epsilon_n$$

$$= n \left[ \sum_{j=1}^{|\mathcal{U}^c|} I \left( \begin{array}{c} X_{\mathcal{V},Q}, X_{[1:l_j-1]\setminus\{l_1,\ldots,l_{j-1}\},Q}, \\ Y_{[1:l_j-1]\setminus\{l_1,\ldots,l_{j-1}\},Q} \end{array} ; Y_{l_j,Q} \left| \begin{array}{c} X_{\mathcal{V}^c,Q}, Q, \\ X_{l_1,Q}, X_{l_2,Q}, \ldots, X_{l_{j-1},Q}, \\ Y_{l_1,Q}, \ldots, Y_{l_{j-1},Q} \end{array} \right. \right) \right.$$

$$\left. + I(X_{\mathcal{S},Q}, Y_{\mathcal{U},Q}; Y_{\mathcal{V}^c,Q} | Y_{\mathcal{U}^c,Q}, X_{\mathcal{S}^c,Q}, Q) \right] + n\epsilon_n$$

$$\overset{(h)}{\leq} n \left[ \sum_{j=1}^{|\mathcal{U}^c|} I \left( \begin{array}{c} X_{\mathcal{V},Q}, X_{[1:l_j-1]\setminus\{l_1,\ldots,l_{j-1}\},Q}, \\ Y_{[1:l_j-1]\setminus\{l_1,\ldots,l_{j-1}\},Q} \end{array} ; Y_{l_j,Q} \left| \begin{array}{c} X_{\mathcal{V}^c,Q}, \\ X_{l_1,Q}, \ldots, X_{l_{j-1},Q}, \\ Y_{l_1,Q}, \ldots, Y_{l_{j-1},Q} \end{array} \right. \right) \right.$$

$$\left. + I(X_{\mathcal{S},Q}, Y_{\mathcal{U},Q}; Y_{\mathcal{V}^c,Q} | Y_{\mathcal{U}^c,Q}, X_{\mathcal{S}^c,Q}) \right] + n\epsilon_n$$

$$\overset{(i)}{=} n \left[ \sum_{j=1}^{|\mathcal{U}^c|} I \left( \begin{array}{c} X_{\mathcal{V}}, X_{[1:l_j-1]\setminus\{l_1,\ldots,l_{j-1}\}}, \\ Y_{[1:l_j-1]\setminus\{l_1,\ldots,l_{j-1}\}} \end{array} ; Y_{l_j} \left| \begin{array}{c} X_{\mathcal{V}^c}, X_{l_1}, \ldots, X_{l_{j-1}}, \\ Y_{l_1}, \ldots, Y_{l_{j-1}} \end{array} \right. \right) \right.$$

$$\left. + I(X_{\mathcal{S}}, Y_{\mathcal{U}}; Y_{\mathcal{V}^c} | Y_{\mathcal{U}^c}, X_{\mathcal{S}^c}) \right] + n\epsilon_n,$$

where $\mathcal{M}$ is the set of all messages, $\mathcal{T} = \{(j,k) : j \in \mathcal{S}, k \in \mathcal{S}^c\}$, $Q$ is a time-sharing random variable taking values uniformly in $[1 : n]$ and independent of other variables, $(a)$ is because messages are independent, $(b)$ is because of Fano's inequality, $(c)$ is because $Y_{\mathcal{S}^c,i} = (Y_{\mathcal{U}^c,i}, Y_{\mathcal{V}^c,i})$, $(d)$ and $(e)$ are because $X_{\mathcal{V}^c,i}$ is a function of $(M_{\mathcal{T}^c}, Y_{\mathcal{V}^c}^{i-1})$ and $X_{\mathcal{U}^c,i}$ is a function of $(M_{\mathcal{T}^c}, Y_{\mathcal{U}^c}^{i})$, $(f)$ is because conditioning reduces entropy, $(g)$ is because of Markov chains $(\mathcal{M}, Y_{\mathcal{S}^c}^{i-1}) \to (X_{\mathcal{V},i}, X_{\mathcal{V}^c,i}, X_{[1:l_j-1],i}, Y_{[1:l_j-1],i}) \to Y_{l_j,i}$ and $(\mathcal{M}, Y_{\mathcal{S}^c}^{i-1})$
$\to (X_{\mathcal{S},i}, X_{\mathcal{S}^c,i}, Y_{\mathcal{U},i}, Y_{\mathcal{U}^c,i}) \to Y_{\mathcal{V}^c,i}$, $(h)$ is because of Markov chains
$Q - (X_{\mathcal{V}^c,Q}, X_{\mathcal{V},Q}, X_{[1:l_j-1],Q}, Y_{[1:l_j-1],Q}) - Y_{l_j,Q}$ and $Q - (Y_{\mathcal{U},Q}, Y_{\mathcal{U}^c,Q}, X_{\mathcal{S},Q}, X_{\mathcal{S}^c,Q}) - Y_{l_j,Q}$, and
$(i)$ is by defining $X = X_Q$, $Y = Y_Q$.

## Appendix B

## Proof of Theorem 2

For some $\epsilon_n \to 0$ as $n \to \infty$, we get

$$n \sum_{j \in \mathcal{V}, k \in \mathcal{S}^c} R_{jk}$$

$$= \sum_{j \in \mathcal{V}, k \in \mathcal{S}^c} H(M_{jk})$$

$$= H(M_{\mathcal{T}} | M_{\mathcal{T}^c})$$

$$= I(M_{\mathcal{T}}; Y_{\mathcal{S}^c}^n | M_{\mathcal{T}^c}) + H(M_{\mathcal{T}} | Y_{\mathcal{S}^c}^n, M_{\mathcal{T}^c})$$

$$\leq \sum_{i=1}^n I(M_{\mathcal{T}}; Y_{\mathcal{S}^c, i} | M_{\mathcal{T}^c}, Y_{\mathcal{S}^c}^{i-1}) + n\epsilon_n$$

$$= \sum_{i=1}^n \big[ H(Y_{\mathcal{S}^c, i} | M_{\mathcal{T}^c}, Y_{\mathcal{S}^c}^{i-1})$$

$$- H(Y_{\mathcal{S}^c, i} | \mathcal{M}, Y_{\mathcal{S}^c}^{i-1}) \big] + n\epsilon_n$$

$$= \sum_{i=1}^n \big[ H(Y_{\mathcal{U}^c, i} | M_{\mathcal{T}^c}, Y_{\mathcal{S}^c}^{i-1})$$

$$- H(Y_{\mathcal{U}^c, i} | \mathcal{M}, Y_{\mathcal{S}^c}^{i-1})$$

$$+ H(Y_{\mathcal{V}^c, i} | M_{\mathcal{T}^c}, Y_{\mathcal{S}^c}^{i-1}, Y_{\mathcal{U}^c, i})$$

$$- H(Y_{\mathcal{V}^c, i} | \mathcal{M}, Y_{\mathcal{S}^c}^{i-1}, Y_{\mathcal{U}^c, i}) \big] + n\epsilon_n$$

$$\overset{(a)}{=} \sum_{i=1}^n \big[ H(Y_{\mathcal{U}^c, i} | M_{\mathcal{T}^c}, Y_{\mathcal{S}^c}^{i-1}, X_{\mathcal{V}^c, i})$$

$$- H(Y_{\mathcal{U}^c, i} | \mathcal{M}, Y_{\mathcal{S}^c}^{i-1}, X_{\mathcal{V}^c, i})$$

$$+ H(Y_{\mathcal{V}^c, i} | M_{\mathcal{T}^c}, Y_{\mathcal{S}^c}^{i-1}, Y_{\mathcal{U}^c, i}, X_{\mathcal{V}^c, i})$$

$$- H(Y_{\mathcal{V}^c, i} | \mathcal{M}, Y_{\mathcal{S}^c}^{i-1}, Y_{\mathcal{U}^c, i}, X_{\mathcal{V}^c, i}) \big] + n\epsilon_n$$

$$\begin{aligned}
&= \sum_{i=1}^{n} \left[ \sum_{j=1}^{|\mathcal{U}^c|} \left\{ H(Y_{l_j,i}|M_{\mathcal{T}^c}, Y_{\mathcal{S}^c}^{i-1}, X_{\mathcal{V}^c,i}, Y_{l_1,i}, \ldots, Y_{l_{j-1},i}, X_{l_1,i}, \ldots, X_{l_{j-1},i}) \right. \right. \\
&\quad \left. - H(Y_{l_j,i}|\mathcal{M}, Y_{\mathcal{S}^c}^{i-1}, X_{\mathcal{V}^c,i}, Y_{l_1,i}, \ldots, Y_{l_{j-1},i}, X_{l_1,i}, \ldots, X_{l_{j-1},i}) \right\} \\
&\quad + H(Y_{\mathcal{V}^c,i}|M_{\mathcal{T}^c}, Y_{\mathcal{S}^c}^{i-1}, Y_{\mathcal{U}^c,i}, X_{\mathcal{V}^c,i}) \\
&\quad \left. - H(Y_{\mathcal{V}^c,i}|\mathcal{M}, Y_{\mathcal{S}^c}^{i-1}, Y_{\mathcal{U}^c,i}, X_{\mathcal{V}^c,i}) \right] + n\epsilon_n \\
&\overset{(b)}{\leq} \sum_{i=1}^{n} \left[ \sum_{j=1}^{|\mathcal{U}^c|} \left\{ H(Y_{l_j,i}|Y_{\mathcal{U}^c}^{i-1}, X_{\mathcal{V}^c,i}, Y_{l_1,i}, \ldots, Y_{l_{j-1},i}, X_{l_1,i}, \ldots, X_{l_{j-1},i}) \right. \right. \\
&\quad \left. - H(Y_{l_j,i}|\mathcal{M}, Y_{\mathcal{S}^c}^{i-1}, X_{\mathcal{V}^c,i}, Y_{[1:l_j-1],i}, X_{[1:l_j-1],i}, X_{\mathcal{V},i}) \right\} \\
&\quad + H(Y_{\mathcal{V}^c,i}|Y_{\mathcal{U}^c}^{i-1}, Y_{\mathcal{U}^c,i}, X_{\mathcal{V}^c,i}) \\
&\quad \left. - H(Y_{\mathcal{V}^c,i}|\mathcal{M}, Y_{\mathcal{S}^c}^{i-1}, Y_{\mathcal{U}^c,i}, X_{\mathcal{V}^c,i}, Y_{\mathcal{U}}^{i-1}, X_{\mathcal{V},i}) \right] + n\epsilon_n \\
&\overset{(c)}{=} \sum_{i=1}^{n} \left[ \sum_{j=1}^{|\mathcal{U}^c|} \left\{ H(Y_{l_j,i}|Y_{\mathcal{U}^c}^{i-1}, X_{\mathcal{V}^c,i}, Y_{l_1,i}, \ldots, Y_{l_{j-1},i}, X_{l_1,i}, \ldots, X_{l_{j-1},i}) \right. \right. \\
&\quad \left. - H(Y_{l_j,i}|Y_{\mathcal{U}^c}^{i-1}, X_{\mathcal{V}^c,i}, Y_{[1:l_j-1],i}, X_{[1:l_j-1],i}, X_{\mathcal{V},i}) \right\} \\
&\quad + H(Y_{\mathcal{V}^c,i}|Y_{\mathcal{U}^c}^{i-1}, Y_{\mathcal{U}^c,i}, X_{\mathcal{V}^c,i}) \\
&\quad \left. - H(Y_{\mathcal{V}^c,i}|Y_{\mathcal{U}^c}^{i-1}, Y_{\mathcal{U}^c,i}, X_{\mathcal{V}^c,i}, Y_{\mathcal{U}}^{i-1}, X_{\mathcal{V},i}) \right] + n\epsilon_n \\
&= \sum_{i=1}^{n} \left[ \sum_{j=1}^{|\mathcal{U}^c|} I \left( \begin{array}{c} X_{\mathcal{V},i}, X_{[1:l_j-1]\setminus\{l_1,\ldots,l_{j-1}\},i}, \\ Y_{[1:l_j-1]\setminus\{l_1,\ldots,l_{j-1}\},i} \end{array} ; Y_{l_j,i} \left| \begin{array}{c} Y_{\mathcal{U}^c}^{i-1}, X_{\mathcal{V}^c,i}, \\ X_{l_1,i}, \ldots, X_{l_{j-1},i}, \\ Y_{l_1,i}, \ldots, Y_{l_{j-1},i} \end{array} \right. \right) \right. \\
&\quad \left. + I(X_{\mathcal{V},i}, Y_{\mathcal{U}}^{i-1}; Y_{\mathcal{V}^c,i}|Y_{\mathcal{U}^c}^{i-1}, Y_{\mathcal{U}^c,i}, X_{\mathcal{V}^c,i}) \right] + n\epsilon_n \\
&\overset{(d)}{=} \sum_{i=1}^{n} \left[ \sum_{j=1}^{|\mathcal{U}^c|} I \left( \begin{array}{c} X_{\mathcal{V},i}, X_{[1:l_j-1]\setminus\{l_1,\ldots,l_{j-1}\},i}, \\ Y_{[1:l_j-1]\setminus\{l_1,\ldots,l_{j-1}\},i} \end{array} ; Y_{l_j,i} \left| \begin{array}{c} U_{\mathcal{U}^c,i}, X_{\mathcal{V}^c,i}, \\ X_{l_1,i}, \ldots, X_{l_{j-1},i}, \\ Y_{l_1,i}, \ldots, Y_{l_{j-1},i} \end{array} \right. \right) \right. \\
&\quad \left. + I(X_{\mathcal{V},i}, U_{\mathcal{U},i}; Y_{\mathcal{V}^c,i}|U_{\mathcal{U}^c,i}, Y_{\mathcal{U}^c,i}, X_{\mathcal{V}^c,i}) \right] + n\epsilon_n
\end{aligned}$$

$$= \sum_{i=1}^{n} \left[ \sum_{j=1}^{|\mathcal{U}^c|} I \left( \begin{array}{c} X_{\mathcal{V},i}, X_{[1:l_j-1]\setminus\{l_1,\ldots,l_{j-1}\},i} \\ Y_{[1:l_j-1]\setminus\{l_1,\ldots,l_{j-1}\},i} \end{array} ; Y_{l_j,i} \middle| \begin{array}{c} U_{\mathcal{U}^c,i}, X_{\mathcal{V}^c,i}, \\ X_{l_1,i},\ldots,X_{l_{j-1},i}, \\ Y_{l_1,i},\ldots,Y_{l_{j-1},i}, Q=i \end{array} \right) \right.$$

$$\left. + I(X_{\mathcal{V},i}, U_{\mathcal{U},i}; Y_{\mathcal{V}^c,i} | U_{\mathcal{U}^c,i}, Y_{\mathcal{U}^c,i}, X_{\mathcal{V}^c,i}, Q=i) \right] + n\epsilon_n$$

$$= n \left[ \sum_{j=1}^{|\mathcal{U}^c|} I \left( \begin{array}{c} X_{\mathcal{V},Q}, X_{[1:l_j-1]\setminus\{l_1,\ldots,l_{j-1}\},Q} \\ Y_{[1:l_j-1]\setminus\{l_1,\ldots,l_{j-1}\},Q} \end{array} ; Y_{l_j,Q} \middle| \begin{array}{c} U_{\mathcal{U}^c,Q}, X_{\mathcal{V}^c,Q}, \\ X_{l_1,Q},\ldots,X_{l_{j-1},Q}, \\ Y_{l_1,Q},\ldots,Y_{l_{j-1},Q}, Q \end{array} \right) \right.$$

$$\left. + I(X_{\mathcal{V},Q}, U_{\mathcal{U},Q}; Y_{\mathcal{V}^c,Q} | U_{\mathcal{U}^c,Q}, Y_{\mathcal{U}^c,Q}, X_{\mathcal{V}^c,Q}, Q) \right] + n\epsilon_n$$

$$\overset{(e)}{\leq} n \left[ \sum_{j=1}^{|\mathcal{U}^c|} I \left( \begin{array}{c} X_{\mathcal{V},Q}, X_{[1:l_j-1]\setminus\{l_1,\ldots,l_{j-1}\},Q} \\ Y_{[1:l_j-1]\setminus\{l_1,\ldots,l_{j-1}\},Q} \end{array} ; Y_{l_j,Q} \middle| \begin{array}{c} U_{\mathcal{U}^c,Q}, X_{\mathcal{V}^c,Q}, \\ X_{l_1,Q},\ldots,X_{l_{j-1}}, \\ Y_{l_1,Q},\ldots,Y_{l_{j-1},Q} \end{array} \right) \right.$$

$$\left. + I(X_{\mathcal{V},Q}, U_{\mathcal{U},Q}; Y_{\mathcal{V}^c,Q} | U_{\mathcal{U}^c,Q}, Y_{\mathcal{U}^c,Q}, X_{\mathcal{V}^c,Q}) \right] + n\epsilon_n$$

$$\overset{(f)}{=} n \left[ \sum_{j=1}^{|\mathcal{U}^c|} I \left( \begin{array}{c} X_{\mathcal{V}}, X_{[1:l_j-1]\setminus\{l_1,\ldots,l_{j-1}\}}, \\ Y_{[1:l_j-1]\setminus\{l_1,\ldots,l_{j-1}\}} \end{array} ; Y_{l_j} \middle| \begin{array}{c} U_{\mathcal{U}^c}, X_{\mathcal{V}^c}, \\ X_{l_1},\ldots,X_{l_{j-1}}, \\ Y_{l_1},\ldots,Y_{l_{j-1}} \end{array} \right) \right.$$

$$\left. + I(X_{\mathcal{V}}, U_{\mathcal{U}}; Y_{\mathcal{V}^c} | U_{\mathcal{U}^c}, Y_{\mathcal{U}^c}, X_{\mathcal{V}^c}) \right] + n\epsilon_n,$$

where $\mathcal{M}$ is the set of all messages, $\mathcal{T} = \{(j,k) : j \in \mathcal{S}, k \in \mathcal{S}^c\}$, $Q$ is a time-sharing random variable taking values uniformly in $[1:n]$ and independent of other variables, $(a)$ is because $X_{\mathcal{V}^c,i}$ is a function of $(M_{\mathcal{T}^c}, Y_{\mathcal{V}^c}^{i-1})$, $(b)$ is because conditioning reduces entropy, $(c)$ is because of Markov chains $(\mathcal{M}, Y_{\mathcal{V}^c}^{i-1}) \to (Y_{\mathcal{U}^c}^{i-1}, X_{\mathcal{V}^c,i}, X_{\mathcal{V},i}, Y_{1[1:l_j-1],i}, X_{[1:l_j-1],i}) \to Y_{l_j,i}$ and $(\mathcal{M}, Y_{\mathcal{V}^c}^{i-1}) \to (Y_{\mathcal{U}^c,i}^{i-1}, Y_{\mathcal{U}^c,i}, X_{\mathcal{V}^c,i}, Y_{\mathcal{U}}^{i-1}, X_{\mathcal{V},i}) \to Y_{\mathcal{V}^c,i}$, where the latter is from Lemma 1, $(d)$ is by defining $U_i = Y^{i-1}$, $(e)$ is because of Markov chains

$Q \to (U_{\mathcal{U}^c,Q}, X_{\mathcal{V}^c,Q}, X_{\mathcal{V},Q}, X_{[1:l_j-1],Q}, Y_{[1:l_j-1],Q}) \to Y_{l_j,Q}$ and

$Q \to (U_{\mathcal{U}^c,Q}, Y_{\mathcal{U}^c,Q}, X_{\mathcal{V}^c,Q}, X_{\mathcal{V},Q}, U_{\mathcal{U},Q}) \to Y_{\mathcal{V}^c,Q}$, where the latter is from Lemma 2, and $(f)$ is by defining $X = X_Q$, $Y = Y_Q$, $U = U_Q$.

*Lemma 1:* The variables $\mathcal{M}, X_{\mathcal{V},i}, X_{\mathcal{V}^c,i}, Y_{\mathcal{U},i}, Y_{\mathcal{U}^c,i}, Y_{\mathcal{V},i}, Y_{\mathcal{V}^c,i}, Y_{\mathcal{U}}^{i-1}, Y_{\mathcal{U}^c}^{i-1}, Y_{\mathcal{V}^c}^{i-1}$ form a Markov chain as follows, where $i \in [1:n], \mathcal{U} = \mathcal{S} \cap \mathcal{N}_0, \mathcal{U}^c = \mathcal{N}_0 \setminus \mathcal{U} = \{l_1,\ldots,l_{|\mathcal{U}^c|}\}, l_k < l_j$ if $k < j$,

$\mathcal{V} = \mathcal{S} \cap \mathcal{N}_1$, and $\mathcal{V}^c = \mathcal{N}_1 \setminus \mathcal{V}$ for $\mathcal{S} \subset [1:K]$.

$$(\mathcal{M}, Y_{\mathcal{V}^c}^{i-1}) \rightarrow (X_{\mathcal{V},i}, X_{\mathcal{V}^c,i}, Y_{\mathcal{U}}^{i-1}, Y_{\mathcal{U}^c}^{i-1}) \rightarrow (Y_{\mathcal{U},i}, Y_{\mathcal{U}^c,i}, Y_{\mathcal{V},i}, Y_{\mathcal{V}^c,i})$$

*Proof:* From the induced distribution

$$p(m_{\mathcal{N}_1}, x_1^n, \ldots, x_K^n, y_1^n, \ldots, y_K^n)$$
$$= \left( \prod_{j \in \mathcal{N}_1} \prod_{k=1}^{K} p(m_{jk}) \right) \prod_{i=1}^{n} \left[ \left( \prod_{k \in \mathcal{N}_1} p(x_{ki}|m_{k1}, \ldots, m_{kK}, y_k^{i-1}) \right) \right.$$
$$\left( \prod_{j=1}^{|\mathcal{N}_0|} p(y_{ji}|x_{\mathcal{N}_1,i}, x_{[1:j-1],i}, y_{[1:j-1],i}) p(x_{ji}|y_j^i) \right)$$
$$\left. p(y_{\mathcal{N}_1,i}|x_{\mathcal{N}_1,i}, x_{\mathcal{N}_0,i}, y_{\mathcal{N}_0,i}) \right],$$

we get the following relations.

For $j \in \mathcal{N}_0$ and $i \in [1:n]$,

$$I \left( Y_{ji}; \left. \begin{array}{c} \mathcal{M}, X_{\mathcal{U}}^{i-1}, X_{\mathcal{U}^c}^{i-1}, \\ X_{\mathcal{V}}^{i-1}, X_{\mathcal{V}^c}^{i-1}, \\ Y_{\mathcal{U}}^{i-1}, Y_{\mathcal{U}^c}^{i-1}, \\ Y_{\mathcal{V}}^{i-1}, Y_{\mathcal{V}^c}^{i-1} \end{array} \right| \begin{array}{c} X_{\mathcal{N}_1,i} \\ X_{[1:j-1],i}, \\ Y_{[1:j-1],i} \end{array} \right)$$

$$= I \left( Y_{ji}; \left. \begin{array}{c} \mathcal{M}, X_{\mathcal{U}}^{i-1}, X_{\mathcal{U}^c}^{i-1}, \\ X_{\mathcal{V}}^{i-1}, X_{\mathcal{V}^c}^{i-1}, \\ Y_{\mathcal{U}}^{i-1}, Y_{\mathcal{U}^c}^{i-1}, \\ Y_{\mathcal{V}}^{i-1}, Y_{\mathcal{V}^c}^{i-1} \end{array} \right| \begin{array}{c} X_{\mathcal{V},i}, X_{\mathcal{V}^c,i}, \\ X_{[1:j-1],i}, Y_{[1:j-1],i} \end{array} \right)$$

$$= 0,$$

and

$$I\left(\begin{array}{c} \mathcal{M}, X_{\mathcal{U}}^{i-1}, X_{\mathcal{U}^c}^{i-1}, \\ X_{\mathcal{V}}^{i-1}, X_{\mathcal{V}^c}^{i-1}, \\ Y_{\mathcal{U}}^{i-1}, Y_{\mathcal{U}^c}^{i-1}, \\ Y_{\mathcal{V}}^{i-1}, Y_{\mathcal{V}^c}^{i-1} \end{array} \middle| X_{\mathcal{N}_1,i}, X_{\mathcal{N}_0,i}, Y_{\mathcal{N}_0,i}\right)$$

$$= I\left(Y_{\mathcal{V},i}, Y_{\mathcal{V}^c,i}; \begin{array}{c} \mathcal{M}, X_{\mathcal{U}}^{i-1}, X_{\mathcal{U}^c}^{i-1}, \\ X_{\mathcal{V}}^{i-1}, X_{\mathcal{V}^c}^{i-1}, \\ Y_{\mathcal{U}}^{i-1}, Y_{\mathcal{U}^c}^{i-1}, \\ Y_{\mathcal{V}}^{i-1}, Y_{\mathcal{V}^c}^{i-1} \end{array} \middle| \begin{array}{c} X_{\mathcal{U},i}, X_{\mathcal{U}^c,i}, \\ X_{\mathcal{V},i}, X_{\mathcal{V}^c,i}, \\ Y_{\mathcal{U},i}, Y_{\mathcal{U}^c,i} \end{array}\right)$$

$$= 0,$$

where $\mathcal{N}_0 = \mathcal{U} \cup \mathcal{U}^c$ and $\mathcal{N}_1 = \mathcal{V} \cup \mathcal{V}^c$. From above mutual informations, we get for $j \in \mathcal{N}_0$ and $i \in [1:n]$,

$$I\left(Y_{j,i}; \mathcal{M}, Y_{\mathcal{V}^c}^{i-1} | X_{\mathcal{V},i}, X_{\mathcal{V}^c,i}, Y_{\mathcal{U}}^{i-1}, Y_{\mathcal{U}^c}^{i-1}, Y_{[1:j-1],i}\right)$$

$$= I\left(Y_{j,i}; \mathcal{M}, Y_{\mathcal{V}^c}^{i-1} \middle| \begin{array}{c} X_{\mathcal{V},i}, X_{\mathcal{V}^c,i}, Y_{\mathcal{U}}^{i-1}, Y_{\mathcal{U}^c}^{i-1} \\ X_{[1:j-1],i}, Y_{[1:j-1],i} \end{array}\right) \tag{14}$$

$$= 0,$$

and

$$I\left(Y_{\mathcal{V},i}, Y_{\mathcal{V}^c,i}; \mathcal{M}, Y_{\mathcal{V}^c}^{i-1} \middle| \begin{array}{c} X_{\mathcal{V},i}, X_{\mathcal{V}^c,i}, \\ Y_{\mathcal{U},i}, Y_{\mathcal{U}^c,i}, \\ Y_{\mathcal{U}}^{i-1}, Y_{\mathcal{U}^c}^{i-1} \end{array}\right)$$

$$= I\left(Y_{\mathcal{V},i}, Y_{\mathcal{V}^c,i}; \mathcal{M}, Y_{\mathcal{V}^c}^{i-1} \middle| \begin{array}{c} X_{\mathcal{U},i}, X_{\mathcal{U}^c,i}, \\ X_{\mathcal{V},i}, X_{\mathcal{V}^c,i}, \\ Y_{\mathcal{U},i}, Y_{\mathcal{U}^c,i} \end{array}\right) \tag{15}$$

$$= 0,$$

where the first equalities both in (14) and (15) are because $X_{ji} \in \mathcal{N}_0$ is a function of $Y_j^i \in \mathcal{N}_0$.

By summing over $j \in \mathcal{N}_0 = \mathcal{U} \cup \mathcal{U}^c$, (14) becomes

$$
\begin{aligned}
\sum_{j=1}^{|\mathcal{N}_0|} & I\left(Y_{j,i}; \mathcal{M}, Y_{\mathcal{V}^c}^{i-1} \;\middle|\; \begin{matrix} X_{\mathcal{V},i}, X_{\mathcal{V}^c,i}, Y_{\mathcal{U}}^{i-1}, Y_{\mathcal{U}^c}^{i-1}, \\ Y_{[1:j-1],i} \end{matrix} \right) \\
&= I\left(Y_{\mathcal{N}_0,i}; \mathcal{M}, Y_{\mathcal{V}^c}^{i-1} \;\middle|\; X_{\mathcal{V},i}, X_{\mathcal{V}^c,i}, Y_{\mathcal{U}}^{i-1}, Y_{\mathcal{U}^c}^{i-1} \right) \\
&= I\left(Y_{\mathcal{U},i}, Y_{\mathcal{U}^c,i}; \mathcal{M}, Y_{\mathcal{V}^c}^{i-1} \;\middle|\; X_{\mathcal{V},i}, X_{\mathcal{V}^c,i}, Y_{\mathcal{U}}^{i-1}, Y_{\mathcal{U}^c}^{i-1} \right) \\
&= 0
\end{aligned}
\tag{16}
$$

By combining (15) and (16), we get

$$
I\left(Y_{\mathcal{U},i}, Y_{\mathcal{U}^c,i}, Y_{\mathcal{V},i}, Y_{\mathcal{V}^c,i}; \mathcal{M}, Y_{\mathcal{V}^c}^{i-1} \;\middle|\; \begin{matrix} X_{\mathcal{V},i}, X_{\mathcal{V}^c,i}, \\ Y_{\mathcal{U}}^{i-1}, Y_{\mathcal{U}^c}^{i-1} \end{matrix} \right) = 0.
$$

■

*Lemma 2:* The variables $Q, X_{\mathcal{V},Q}, X_{\mathcal{V}^c,Q}, Y_{\mathcal{U},Q}, Y_{\mathcal{U}^c,Q}, Y_{\mathcal{V},Q}, Y_{\mathcal{V}^c,Q}, U_{\mathcal{U},Q}, U_{\mathcal{U}^c,Q}$ form a Markov chain as follows, where $Q$ is a time-sharing random variable taking values uniformly in $[1:n]$, $\mathcal{U} = \mathcal{S} \cap \mathcal{N}_0$, $\mathcal{U}^c = \mathcal{N}_0 \setminus \mathcal{U} = \{l_1, \ldots, l_{|\mathcal{U}^c|}\}$, $l_k < l_j$ if $k < j$, $\mathcal{V} = \mathcal{S} \cap \mathcal{N}_1$, and $\mathcal{V}^c = \mathcal{N}_1 \setminus \mathcal{V}$ for $\mathcal{S} \subset [1:K]$.

$$
Q \rightarrow (X_{\mathcal{V},Q}, X_{\mathcal{V}^c,Q}, U_{\mathcal{U},Q}, U_{\mathcal{U}^c,Q}) \rightarrow (Y_{\mathcal{U},Q}, Y_{\mathcal{U}^c,Q}, Y_{\mathcal{V},Q}, Y_{\mathcal{V}^c,Q})
$$

*Proof:* Similarly as Lemma 1, the following relations hold for $j \in \mathcal{N}_0$ and $i \in [1:n]$.

$$
I\left(Y_{j,Q}; \begin{matrix} Q, X_{\mathcal{U}}^{Q-1}, X_{\mathcal{U}^c}^{Q-1}, \\ X_{\mathcal{V}}^{Q-1}, X_{\mathcal{V}^c}^{Q-1}, \\ Y_{\mathcal{U}}^{Q-1}, Y_{\mathcal{U}^c}^{Q-1}, \\ Y_{\mathcal{V}}^{Q-1}, Y_{\mathcal{V}^c}^{Q-1} \end{matrix} \;\middle|\; \begin{matrix} X_{\mathcal{V},Q}, X_{\mathcal{V}^c,Q}, \\ X_{[1:j-1],Q}, \\ Y_{[1:j-1],Q} \end{matrix} \right) = 0
$$

$$
I\left(Y_{\mathcal{V},Q}, Y_{\mathcal{V}^c,Q}; \begin{matrix} \mathcal{M}, Q, X_{\mathcal{U}}^{Q-1}, X_{\mathcal{U}^c}^{Q-1}, \\ X_{\mathcal{V}}^{Q-1}, X_{\mathcal{V}^c}^{Q-1}, \\ Y_{\mathcal{U}}^{Q-1}, Y_{\mathcal{U}^c}^{Q-1}, \\ Y_{\mathcal{V}}^{Q-1}, Y_{\mathcal{V}^c}^{Q-1} \end{matrix} \;\middle|\; \begin{matrix} X_{\mathcal{U},Q}, X_{\mathcal{U}^c,Q}, \\ X_{\mathcal{V},Q}, X_{\mathcal{V}^c,Q}, \\ Y_{\mathcal{U},Q}, Y_{\mathcal{U}^c,Q} \end{matrix} \right) = 0.
$$

From above mutual informations, we get

$$I\left(Y_{j,Q}; Q | X_{\mathcal{V},Q}, X_{\mathcal{V}^c,Q}, Y_{\mathcal{U}}^{Q-1}, Y_{\mathcal{U}^c}^{Q-1}, Y_{[1:j-1],Q}\right)$$

$$= I\left(Y_{j,Q}; Q \left| \begin{array}{c} X_{\mathcal{V},Q}, X_{\mathcal{V}^c,Q}, Y_{\mathcal{U}}^{Q-1}, Y_{\mathcal{U}^c}^{Q-1} \\ X_{[1:j-1],Q}, Y_{[1:j-1],Q} \end{array} \right.\right) \tag{17}$$

$$= 0,$$

and

$$I\left(Y_{\mathcal{V},Q}, Y_{\mathcal{V}^c,Q}; Q \left| \begin{array}{c} X_{\mathcal{V},Q}, X_{\mathcal{V}^c,Q}, \\ Y_{\mathcal{U},Q}, Y_{\mathcal{U}^c,Q}, \\ Y_{\mathcal{U}}^{Q-1}, Y_{\mathcal{U}^c}^{Q-1} \end{array} \right.\right)$$

$$= I\left(Y_{\mathcal{V},Q}, Y_{\mathcal{V}^c,Q}; Q \left| \begin{array}{c} X_{\mathcal{U},Q}, X_{\mathcal{U}^c,Q}, \\ X_{\mathcal{V},Q}, X_{\mathcal{V}^c,Q}, \\ Y_{\mathcal{U},Q}, Y_{\mathcal{U}^c,Q} \end{array} \right.\right) \tag{18}$$

$$= 0,$$

where the first equalities both in (17) and (18) are because $X_{j,Q} \in \mathcal{N}_0$ is a function of $Y_j^Q \in \mathcal{N}_0$. By summing over $j \in \mathcal{N}_0 = \mathcal{U} \cup \mathcal{U}^c$, (17) becomes

$$\sum_{j=1}^{|\mathcal{N}_0|} I\left(Y_{j,Q}; Q | X_{\mathcal{V},Q}, X_{\mathcal{V}^c,Q}, Y_{\mathcal{U}}^{Q-1}, Y_{\mathcal{U}^c}^{Q-1}, Y_{[1:j-1],Q}\right)$$

$$= I\left(Y_{\mathcal{N}_0,Q}; Q \left| X_{\mathcal{V},Q}, X_{\mathcal{V}^c,Q}, Y_{\mathcal{U}}^{Q-1}, Y_{\mathcal{U}^c}^{Q-1} \right.\right) \tag{19}$$

$$= I\left(Y_{\mathcal{U},Q}, Y_{\mathcal{U}^c,Q}; Q \left| X_{\mathcal{V},Q}, X_{\mathcal{V}^c,Q}, Y_{\mathcal{U}}^{Q-1}, Y_{\mathcal{U}^c}^{Q-1} \right.\right)$$

$$= 0$$

By combining (18) and (19) and by defining $U_Q = Y^{Q-1}$, we get

$$I\left(Y_{\mathcal{U},Q}, Y_{\mathcal{U}^c,Q}, Y_{\mathcal{V},Q}, Y_{\mathcal{V}^c,Q}; Q \left| \begin{array}{c} X_{\mathcal{V},Q}, X_{\mathcal{V}^c,Q}, \\ U_{\mathcal{U},Q}, U_{\mathcal{U}^c,Q} \end{array} \right.\right) = 0.$$

■

## References

[1] E. C. van der Meulen, "Transmission of information in a t-terminal discrete memoryless channel," Ph. D. dissertation, Univ. of California, Berkely, CA, 1968.

 

[2] ——, "Three-terminal communications channels," *Advances in Applied Probability*, vol. 3, pp. 120–154, 1971.

[3] T. M. Cover and A. El Gamal, "Capacity theorems for the relay channel," *IEEE Trans. Inf. Theory*, vol. 25, pp. 572–584, Sep. 1979.

[4] T. M. Cover and J. A. Thomas, *Elements of information theory*. New York: Wiley, 2006.

[5] A. El Gamal, N. Hassanpour, and J. Mammen, "Relay networks with delays," *IEEE Trans. Inf. Theory*, vol. 53, pp. 3413–3431, Oct. 2007.

[6] L. Wang and M. Naghshvar, "On the capacity of the noncausal relay channel," in *IEEE Int. Symp. Inform. Theory (ISIT)*, St. Petersburg, Russia, 2011.

[7] H. Chang and S.-Y. Chung, "Capacity of strong and very strong Gaussian interference relay-without-delay channels," submitted to IEEE Transactions of Information Theory. [Online]. Available: arXiv:1108.2846/cs.IT 2011.

[8] I.-J. Baik and S.-Y. Chung, "Causal relay networks and new cut-set bounds," in *Proc. 49th Annual Allerton Conference on Communication, Control, and Computing*, Allerton House, Monticello, IL, 2011.

[9] ——, "Causal relay networks with causal side information," in *IEEE Int. Symp. Inform. Theory (ISIT)*, Boston, USA, 2012.

[10] S. L. Fong, R. W. Yeung, and G. Kramer, "Cut-set bound for generalized networks," in *IEEE Int. Symp. Inform. Theory (ISIT)*, Boston, USA, 2012.

[11] G. Kramer, "Networks with in-block memory," submitted to Information Theory Workshop. [Online]. Available: arXiv:1206.5389/cs.IT 2012.